# Optimal demand-responsive connector design: Comparing fully-flexible routing and semi-flexible routing strategies

Li Zhen [a], Weihua Gu [a*]

[a] Department of Electrical and Electronic Engineering, The Hong Kong Polytechnic University, Hong Kong SAR, China

## Abstract

Demand-responsive connector (DRC) services are increasingly recognized for their convenience, comfort, and efficiency, offering seamless integrations between travelers' origins/destinations and major transportation hubs such as rail stations. Past analytical models for DRC optimization often failed to distinguish between two commonly used DRC operating strategies: (i) the "fully-flexible routing" strategy, where a vehicle serves only the requests received before its dispatch through an optimal tour, and (ii) the "semi-flexible routing" strategy, where a vehicle follows a predefined path through a swath to serve requests received en route. Additionally, these models often adopted oversimplified approaches for estimating local tour lengths and capturing the stochastic nature of demand.

This paper distinctly identifies and analyzes the two DRC operating strategies, developing analytical models for each that accurately incorporate the second-order effects of stochastic demand and utilize refined local tour length formulas. Numerical experiments demonstrate that our models reduce cost estimation errors to within 2% for fully-flexible routing and to 0.25% for semi-flexible routing, a significant improvement over the previous errors of 8-12% and 6.3%, respectively. These enhanced models allow for more precise determination of critical demand densities for selecting between the two DRC strategies and the fixed-route feeder service. Our extensive numerical analysis offers many insights, particularly highlighting the transition from fully-flexible to semi-flexible routing as demand and region size increase, before ultimately shifting to fixed-route service. Additionally, zoning is identified as pivotal in DRC service design, with fully-flexible routing favoring square-shaped zones and semi-flexible routing preferring elongated rectangular zones.

## 1. Introduction

### 1.1 Background and literature review on optimal DRC design

Efficient public transportation systems are critical in fostering sustainable urban mobility amidst the accelerating trend of global urbanization. As metropolitan areas and city clusters continue to expand, rapid rail transit links connecting urban cores with outlying satellite towns and neighboring cities have become increasingly important. China's extensive high-speed rail network, which penetrates numerous cities and towns, exemplifies this trend. However, these rail systems often feature extensive station spacings, thereby emphasizing the necessity for faster and more efficient access modes to bridge the substantial gaps between rail stations. Hence, the development of specialized feeder systems has emerged as a critical area of interest. Such systems aim to bolster mass transit network connectivity and streamline urban and intercity mobility by effectively linking passengers from dispersed origins and destinations to trunk transit terminals (Montenegro et al., 2021).

---

[*] Corresponding author. Email: weihua.gu@polyu.edu.hk

Feeder bus services come in two primary forms: fixed-route and flex-route (Sangveraphunsiri et al., 2022). The latter is also termed demand-responsive connectors, or DRCs.[1] Fixed-route services are more economical serving high-demand urban areas with dense lines and stops and short headways (Petit et al., 2021; Wong et al., 2023). However, passengers must walk to and wait at stops. In contrast, DRC services, though typically more costly to operate (Wang et al., 2023), offer the added convenience of door-to-door service. This feature is particularly appealing to business travelers, executive passengers, and those needing extra assistance, such as the elderly or disabled (Kersting et al., 2021). They excel in areas with lower demand (Wang et al., 2014) and are increasingly enhanced by today's information and communication technologies, such as user-friendly booking apps and real-time updates on vehicle location and status (Liu and Ouyang, 2021; Ma et al., 2019).

The study of DRC design has seen the application of various methodologies, each with its strengths and limitations. Discrete mathematical programming models have been notably used to tailor feeder system designs to specific real-world networks, as seen in the work of Pan et al. (2015). These models are detailed and precise but are constrained by substantial computational complexity, often deemed NP-hard problems. Another approach is the use of simulation, as explored by Li and Quadrifoglio (2010), which offers a dynamic representation of DRC operations and can yield a realistic portrayal of their performance. Despite its practicality, the simulation approach is computationally intensive—perhaps even more so than discrete modeling methods—and may not always provide clear analytical insights.

To address the limitations of the above approaches, the so-called "analytical models" (e.g., Li and Quadrifoglio, 2010) have gained traction for their ability to distill complex systems into essential operating variables. These models stand out for their generality, offering insights into the intricate cause-and-effect relationships between key operating features and the optimal DRC design. Moreover, these models can often be solved efficiently, facilitating the extraction of operational insights that are crucial for enhancing our comprehension and implementation of DRCs in practical scenarios. For example, Quadrifoglio and Li (2009) employed analytical models to derive approximate closed-form expressions for the critical demand density for choosing between fixed-route and flex-route services. Their models, while insightful, were limited to single and two-zone scenarios. Qiu et al. (2014), Zheng et al. (2018a), and Zheng et al. (2019) investigated feeder services where vehicles deviate from a set route on an ad-hoc basis to accommodate real-time demand requests. Zheng et al. (2018b) examined the implications of route deviation, enabling buses to alter their fixed routes for more direct passenger service. Sangveraphunsiri et al. (2022) proposed an innovative hybrid concept known as the "Jitney-lite" strategy, which employs a fixed-route service for the outbound leg (from passengers' origins to a mass transit terminal) and a flex-route service for the inbound leg (from the terminal to passengers' destinations).

The aforementioned studies presuppose that a single vehicle serves the entire service region or that the region is divided into a fixed number of zones with one vehicle allocated per zone. These approaches may lead to impractically long tour lengths and travel times, particularly in larger regions. To address this issue, Chang and Schonfeld (1991) developed continuum approximation models that optimize zone partitioning and vehicle sizes for flex-route feeder services. They also pinpointed the critical demand threshold that necessitates a switch between fixed and flex-route services. Subsequent research conducted by Kim and Schonfeld from 2012 to 2015 introduced a series of improvements to the model: optimizing service headways (Kim and Schonfeld, 2012), integrating mixed fleets with varying vehicle sizes (Kim and Schonfeld, 2013), examining multi-region service coordination (Kim and Schonfeld, 2014), and incorporating elastic demand (Kim and Schonfeld, 2015). More recent follow-up works focused on the stochastic dynamic switching between fixed and flex-route services (Guo et al., 2017), and the derivation of closed-form relationship between optimal headway and zone size (Kim et al., 2019).

---

[1] In the present paper, we employ the terms "flex-route feeder" and "DRC" interchangeably. Notably, however, some studies have described nuanced distinctions between these two terms (see, for example, Li and Quadrifoglio, 2010).

**1.2 Tour length estimation under distinct routing strategies**

A fundamental issue in optimizing DRC systems is the estimation of flex-route vehicle tour lengths. This problem aligns with the classic Traveling Salesman Problem (TSP)—to find the shortest tour length that visits a specified number of locations. While established algorithms or commercial solvers are adequate for resolving specific TSP instances (e.g., the one depicted in Fig. 1a), analytical modeling of DRC services demands a general closed-form expression for the expected TSP tour length, rather than case-specific TSP solutions. Fortunately, an asymptotic tour length formula was provided by Beardwood et al. (1959) and later extended by Stein (1978a, b). This foundational result asserts that the expected tour length for visiting a sufficiently large number of points denoted by $q$ can be estimated as $k^*\sqrt{qA}$, where $A$ is the area of the bounded zone containing the $q$ points, and $k^*$ is a constant.[2] This formula has been extensively adopted in the literature of DRC design.

Determining the value of $k^*$ is essential yet challenging. Unfortunately, the existing research has not yielded satisfactory results for DRC model applications. Most relevant studies reported $k^*$ values ranging from 0.712 to 0.765, obtained under the Euclidean distance metric (Bonomi and Lutton, 1984; Lee and Choi, 1994; Percus and Martin, 1996; Stein, 1978a, b). In the context of transportation, however, the Manhattan metric is more commonly used due to the grid-like nature of urban road networks (e.g., Daganzo, 2010; Fan et al., 2018; Nourbakhsh and Ouyang, 2012; Wu et al., 2020). Regrettably, there are scant estimates of $k^*$ that consider Manhattan geometry. Chakraborti and Chakrabarti (2000) proposed a $k^*$ value of 0.93 for cases where $q \to \infty$. For smaller values of $q$, Yang et al. (2020) demonstrated that $k^*$ *is not a constant but varies depending on $q$ and the service zone's geometry*. They offered a regression-based formula $k^* = \left(1.1055 - 0.008q + 1.0297\frac{S}{q}\right)$, where $S$ stands for the aspect ratio of a rectangular zone. Although their formula exhibits commendable precision for $10 \leq q \leq 100$, our numerical tests indicate that its accuracy decreases significantly for even smaller $q$ values ($2 \leq q \leq 15$), which are typical for DRC vehicles serving low-demand areas. *The observed error in these instances can be over 40%* (see Section 2.2.1).

On the other hand, Daganzo (1984) proposed an efficient heuristic method for tour construction, where a service zone is cut into a swath of width $w_0$, as exemplified in Fig. 1b. Vehicles navigate along the swath, making lateral detours to pick up passengers, a strategy known in some DRC literature as "no-backtracking routing" (e.g., Qiu et al., 2014). We label it *semi-flexible routing* in this paper, as the vehicle's route is fixed barring those lateral detours. This routing heuristic allows for an analytical expression for the expected tour length, dependent on swath width $w_0$. Daganzo demonstrated that the minimum expected tour length is $1.15\sqrt{qA}$ when $w_0$ is set to $w_0^* = \sqrt{\frac{3A}{q}}$. This $k^*$ value of 1.15 has since been widely used in DRC literature to estimate tour lengths (e.g., Chang and Schonfeld, 1991). However, it doesn't account for the fact that $w_0^*$ is often unachievable in practice. For rectangular zones that are typically assumed in DRC models, the zone length or width should be an integer multiple of $w_0$, resulting in a $k^*$ value that is necessarily higher than 1.15 and dependent on the zone's dimensions.

---

[2] This formula applies to both types of tours: closed tours, where the vehicle completes a loop by returning to its starting point (Beardwood et al., 1959), and open tours, where the vehicle finishes at a different location and does not loop back (Stein, 1978a).

Despite the imprecision of $k^*$ values reported in the literature, e.g., 0.93 for the optimal TSP routing strategy and 1.15 for semi-flexible routing, a rough comparison still suggests that an optimized TSP tour is on average 19% shorter than a semi-flexible heuristic, implying the great potential of using the former strategy. *Yet, these strategies impact not only tour lengths but also DRC service operations.* The optimized TSP routing, termed *"fully-flexible routing strategy"* in this paper, necessitates prior knowledge of all service requests before a bus is dispatched. This leads to longer passenger wait times at home, including both the interval from request submission to vehicle dispatch and the time until pick-up. Conversely, the semi-flexible routing strategy allows for on-the-fly pick-ups along the swath, provided those requests occur before the bus reaches their locations. This approach dispatches buses without predetermined stops, thereby substantially decreasing passenger wait times by removing the wait for bus dispatch. *The distinct operational mechanisms of each strategy represent an element that is often overlooked in DRC literature.*

*In summary, fully-flexible routing achieves reduced vehicle tour lengths but extends passenger wait times, whereas semi-flexible routing typically shortens wait times at the expense of longer tours.* Therefore, determining which strategy optimally balances passenger experience and operational costs remains unresolved, motivating the present study.

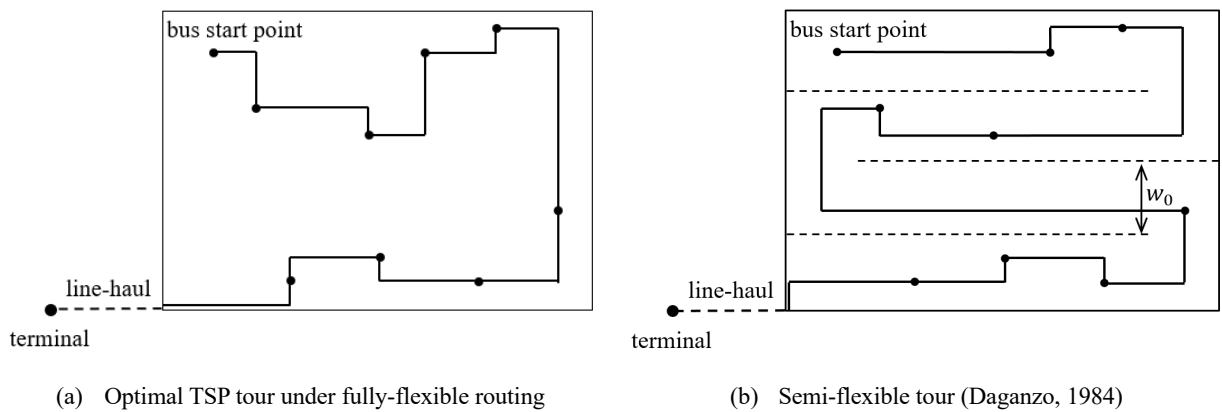

(a) Optimal TSP tour under fully-flexible routing        (b) Semi-flexible tour (Daganzo, 1984)

**Fig. 1. Two routing strategies**

## 1.3 Overview of the paper

This paper addresses the above gaps in the literature and contributes to the research realm of DRC design optimization in the following major ways:

(i) We have precisely defined the fully-flexible and semi-flexible routing strategies for DRC operations and formulated respective optimization models. Through comprehensive numerical analyses, we compared the strategies to determine their relative performance under various operating conditions. Our findings yield key insights instrumental for the optimal DRC service designs and operations.

(ii) To improve the validity of model outcomes, we have recalibrated the estimates of the $k^*$

values for fully-flexible routing in realistic DRC scenarios, which involve smaller occupancies ($q$ values) and diverse service zone dimensions, in the context of the Manhattan distance metric. Furthermore, we have rectified the previously incorrect use of $k^* = 1.15$ for semi-flexible routing by incorporating the constrained relationship between zone dimensions and the swath width, $w_0$.

(iii) Lastly, our research refines and advances the modeling of stochastic demand in DRC design optimization, enhancing modeling accuracy. The necessity of this advancement is underscored by the variability in vehicle occupancy (i.e., the number of passengers a vehicle carries) due to random demand, which persists even under fixed dispatch schedules commonly implemented in DRC services (Wong et al., 2023). *Prior models often relied solely on expected vehicle occupancy, neglecting the variation of occupancy*.[3] This simplification might be adequate for cost functions linear in vehicle occupancy, but it falls short for nonlinear cost components, such as the cumulative passenger wait times and in-vehicle travel times, which we will discuss in Section 2.2.2. This discrepancy arises because *the expected value of a nonlinear function of occupancy is typically not equal to the function's value at the expected occupancy*. Furthermore, the distribution of vehicle occupancy, rather than its expected value, is crucial for determining the correct vehicle size. Hence, using only the expected occupancy for DRC design optimization—a prevalent oversight in the literature—can result in flawed cost estimations and impractical designs. By integrating the vehicle occupancy distribution into our models, we ensure a more nuanced and accurate optimization of DRC design, overcoming limitations of previous research.

In the remainder of this paper, we begin by detailing the problem formulation and developing analytical models optimizing the design of DRC services under the fully-flexible routing and semi-flexible routing strategies (Section 2). Our models take a rigorous approach to accommodating stochastic demand and more accurate estimations of tour lengths. The solution method is elaborated upon in Section 3. In Section 4, we conduct thorough numerical case studies to compare the efficacy of the two routing strategies across a spectrum of operating scenarios. Additionally, we juxtapose the optimally designed DRC services with optimal fixed-route feeder services within various operational contexts. The paper concludes with a summary of findings and a discussion of potential avenues for future research in Section 5.

## 2. Methodology

Assumptions and setup of the DRC design problem are presented in Section 2.1. The cost models under the fully-flexible routing and semi-flexible routing strategies are formulated in Sections 2.2 and 2.3, respectively. Section 2.4 furnishes the optimization formulations under the two strategies. Notations used in these models are summarized in Appendix A.

---

[3] Using expected vehicle occupancy in prior studies (Chang and Schonfeld, 1991; Quadrifoglio and Li, 2009; Kim and Schonfeld, 2012, 2013, 2014, 2015, 2019; Qiu et al., 2014; Zheng et al., 2018a; Zheng et al., 2018b) would be proper if a fixed occupancy was assumed for DRC operations. This assumption may hold for services that allow advance reservations, such as airport shuttle buses that schedule pick-ups well ahead of time. For feeder operations serving commuters, adhering to a fixed occupancy potentially exposes passengers to additional uncertainty regarding vehicle dispatch times when placing their requests. It is also worth noting that such an assumption is usually not explicitly stated in these studies. Moreover, this assumption conflicts with Daganzo's semi-flexible routing strategy, which inherently accommodates variable occupancy levels.

## 2.1 Problem setup and assumptions

Fig. 2 illustrates the idealized network for the DRC service. A DRC bus terminal, marked by a black dot, is depicted at the lower-left corner. This terminal acts as a nexus for transfers, connecting passengers to a transportation hub such as an intercity rail station. The terminal's catchment area is assumed to be a rectangle centered around it. Without loss of generality, we focus our design efforts on a quarter of this rectangular area, i.e., a *service region* denoted by $\Omega \equiv \{(x,y)|0 \leq x \leq L, 0 \leq y \leq W\}$; see Fig. 2. Outbound demand requests are modeled as a spatial Poisson process (Sangveraphunsiri et al., 2022) with a uniform intensity of $\lambda_p$ patrons/km²/h. In the inbound direction, the demand requests follow a similar process but at a distinct intensity of $\lambda_d$ patrons/km²/h.

For both routing strategies, the catchment area is partitioned into $M \times N$ identical rectangular zones, each measuring $l \times w$, where $l = \frac{L}{N}$, $w = \frac{W}{M}$. For ease of reference, zones are identified by coordinates $(m,n)$, where $m$ incrementing from bottom to top and $n$ from left to right. It is assumed that transit vehicles follow a fixed-headway schedule for both collecting and distributing passengers, denoted as $H_{p(m,n)}$ and $H_{d(m,n)}$, respectively, for each zone $(m,n)$. Note that fixed-headway dispatching, in contrast to fixed-occupancy, tends to be more favorable among DRC patrons and is more commonly implemented in real-world operations (Dzisi et al., 2022; Wong et al., 2023). Further note that optimal headways may vary between zones as a result of the differences in line-haul distances associated with each zone. Zones requiring longer line-hauls will generally need longer round-trip times and incur higher costs per trip, which necessitates larger headways to balance those costs.

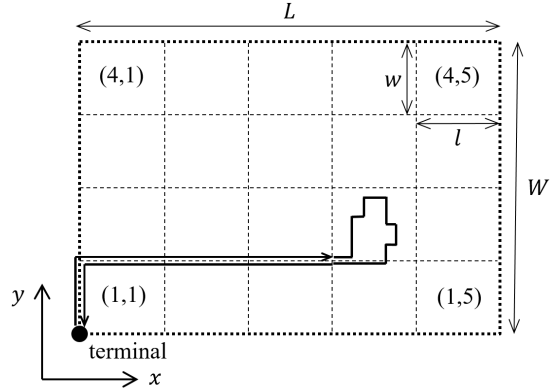

**Fig. 2 DRC service network**

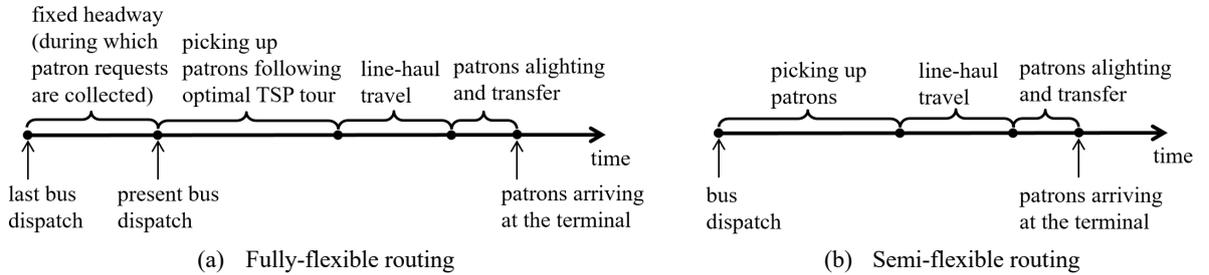

(a) Fully-flexible routing  (b) Semi-flexible routing

**Fig. 3 Service processes of DRC operations in outbound direction**

Fig. 3a demonstrates the service process for the fully-flexible routing strategy, focusing on the outbound direction as an example. Under this strategy, a patron submits a request, perhaps through a mobile phone app, before the next bus dispatch. The bus scheduled to depart will serve all requests received within the headway, following an optimal TSP tour calculated using commercial solvers or established algorithms, such as the Concorde TSP Solver[4]. Subsequently, the bus proceeds on its line-haul journey to the terminal. Under the fixed-headway setting, the number of patrons served by a DRC bus follows a Poisson distribution with respective means $\lambda_p H_{p(m,n)} lw$ for the outbound direction and $\lambda_d H_{d(m,n)} lw$ for the inbound direction, respectively. The variables to be optimized in this strategy include the zoning dimensions, $l$ and $w$ (or, equivalently, $M$ and $N$), the headways, $H_{p(m,n)}$ and $H_{d(m,n)}$, and the bus capacity, $K$.

Contrastingly, the semi-flexible routing strategy depicted in Fig. 3b does not necessitate patron pre-registration prior to the dispatch of buses. Under this scheme, patrons can request trips spontaneously, and they are subsequently picked up by the next bus that passes by, which could already be in the midst of its tour at the time the request is made. The optimization of this strategy revolves around the following decision variables: the zoning dimensions, $l$ and $w$ (or, equivalently, $M$ and $N$), the headways, $H_{p(m,n)}$ and $H_{d(m,n)}$, the bus capacity, $K$, and the swath width, $w_0$. We assume that buses maintain roughly regular headways at each longitudinal position along the swath. Thus, the number of patrons carried by a bus under this scheme also follows a Poisson distribution with mean $\lambda_p H_{p(m,n)} lw$ for outbound and $\lambda_d H_{d(m,n)} lw$ for inbound.

We also adopt the simplifying assumption that only one patron is picked up or dropped off per local tour stop. This assumption is commonly utilized in transit research (e.g., Quadrifoglio and Li, 2009; Sangveraphunsiri et al., 2022).

In line with established practices in the field (Ouyang et al., 2014; Chen and Nie, 2017), our objective is to minimize the total system cost of the DRC service under each routing strategy. This total system cost, sometimes referred to as "generalized cost" in the literature (Daganzo, 2010; Gu et al., 2016; Mei et al., 2021), encapsulates both the travel time experienced by patrons and the operating costs incurred by the agency. Note that patrons' travel time is measured in hours, while agency costs are in monetary terms. To add up these cost terms with disparate units, we convert the agency's costs into equivalent temporal units using the average patron's value of time, expressed in $/hour (Daganzo, 2010). In addition, bus fares, which represent a monetary transfer from patrons to the transit agency, are omitted from the total cost calculation. While fares are a cost to patrons, they are concurrently revenue for the agency, thereby netting out in the context of system cost assessment.

Subsequent sections will elaborate on the cost models for one hour of DRC operations under the fully-flexible and semi-flexible routing strategies.

**2.2 Cost models of the fully-flexible routing strategy**

*2.2.1 Estimation of expected TSP tour length*

Recall that the expected optimal TSP tour length within a designated area $A$ is known to scale asymptotically with $\sqrt{qA}$, where $q$ is the number of stops. For small values of $q$, the scaling factor, $k^*$, depends on both the number of stops and the geometrical features of the region, such as the aspect ratio for rectangular areas. Unfortunately, there exists a gap in the literature regarding the estimates of $k^*$ for fewer ($\leq 15$) stops in Manhattan geometry; see the discussion in Section 1.2.

Given that the passenger load for a DRC bus is commonly lower than 15, we aim to establish an empirical model for $k^*$ grounded in Monte Carlo simulation outcomes in this section. We execute a series of simulation scenarios that encompass a range of stops from 2 to 15 and various aspect ratios from 1 to 3. (we consider that aspect ratios greater than 3 are uncommon in practice.) For each scenario, we randomly generate and optimally solve 500 TSP instances using an exact algorithm implemented in Matlab R2020a. The experimental setup is detailed in Appendix B. Following these simulations, we compute the average $k^*$ for every scenario, which are used to calibrate the model.

---

[4] https://www.math.uwaterloo.ca/tsp/concorde.html

Through regression analysis, the following model, inspired by the Weibull probability density function, emerges as the best fit for the simulation data:

$$k^* = (\beta_1 S + \beta_2) q^{\beta_3} e^{\beta_4 q^{\beta_5}} \tag{1}$$

where $q$ denotes the number of stops and $S \geq 1$ the aspect ratio. The calibrated coefficient values are tabulated as follows:

**Table 1. Values of coefficients in Eq. (1)**

| $\beta_1$ | $\beta_2$ | $\beta_3$ | $\beta_4$ | $\beta_5$ |
|---|---|---|---|---|
| 0.1102 | 1.4569 | -0.1472 | -2.5508 | -2.6396 |

Eq. (1) provides a refined estimation of the optimal TSP tour lengths pertinent to our feeder system. The mean absolute percentage errors (MAPE) across the specified ranges of $q$ and $S$ are showcased in Fig. 4. For comparison, Fig. 4 also includes the MAPE values based on the model from Yang et al. (2020) and the $k^*$ value proposed by Chakraborti and Chakrabarti (2000); see Section 1.2 for the details of those benchmarks. The depicted results indicate that *the error margins of the models from earlier studies are considerably high, with some scenarios yielding errors surpassing 40%. In stark contrast, our method provides a markedly closer estimate to the true values, maintaining a maximum MAPE under 5%.*

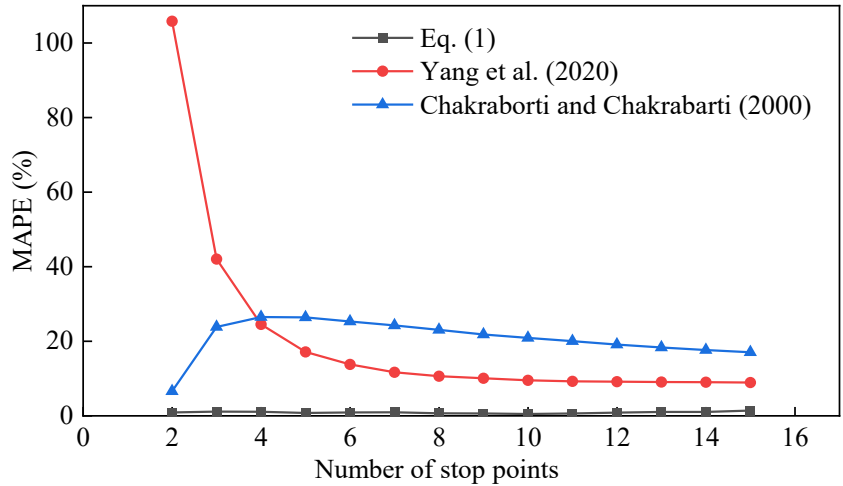

**Fig. 4. The MAPEs of $k^*$ under fully-flexible routing**

*2.2.2 Patrons' travel time cost*

The travel time cost for patrons is broken down into four elements: (i) waiting time at home; (ii) travel time during the local tour; (iii) travel time on the main line-haul; and (iv) transfer time at the terminal. Each of these components is detailed as follows:

(i) <u>Waiting time at home</u>

This element applies exclusively to the outbound leg where patrons are being picked up from home.

The aggregate waiting time at home per operational hour of the service, represented by $C_W$, is calculated by summing the total waiting times for each zone, $C_{W(m,n)}$, where $1 \leq m \leq M$, $1 \leq n \leq N$. A patron's waiting time is comprised of: (a) the interval preceding the bus dispatch; and (b) the duration from bus dispatch until its arrival. The expected value for the first part is $\frac{H_{p(m,n)}}{2}$, since the service requests following a Poisson process are uniformly distributed over time. The second part is, on average, half of the local TSP tour's travel time. Consequently, the total waiting time is expressed as follows:

$$C_{W(m,n)} = \alpha \cdot \left\{ \frac{1}{H_{p(m,n)}} \cdot E\left[Q_{p(m,n)} \cdot \frac{H_{p(m,n)}}{2}\right] + \frac{1}{2H_{p(m,n)}} \cdot E\left[Q_{p(m,n)} \cdot \left(\frac{1}{v_I}(\beta_1 S + \beta_2)(Q_{p(m,n)} + 1)^{\beta_3} e^{\beta_4(Q_{p(m,n)}+1)^{\beta_5}} \sqrt{(Q_{p(m,n)} + 1)lw} + Q_{p(m,n)}\tau_p\right)\right]\right\}$$

$$= \frac{\alpha}{2} E[Q_{p(m,n)}] + \frac{\alpha(\beta_1 S + \beta_2)\sqrt{lw}}{2H_{p(m,n)}v_I} E\left[\left((Q_{p(m,n)} + 1)^{\beta_3 + \frac{3}{2}} - (Q_{p(m,n)} + 1)^{\beta_3 + \frac{1}{2}}\right) e^{\beta_4(Q_{p(m,n)}+1)^{\beta_5}}\right] + \frac{\alpha \tau_p}{2H_{p(m,n)}} E[Q_{p(m,n)}^2] \quad (2)$$

$$C_W = \sum_{n=1}^{N} \sum_{m=1}^{M} C_{W(m,n)} \quad (3)$$

where $E[\cdot]$ denotes the expected value operator; $Q_{p(m,n)}$ the random number of patrons collected by a bus serving zone $(m, n)$; $v_I$ the bus cruising speed; $\beta_1$, $\beta_2$, $\beta_3$, $\beta_4$, $\beta_5$ represent the corresponding coefficients used in Eq. (1); $\tau_p$ denotes the time loss per stop during the outbound tour; and $\alpha \in [0,1]$ is *the discount factor applied to the time value of waiting at home*. This discount factor is introduced to reflect the reduced disutility of waiting at home as opposed to traveling in a vehicle, acknowledging the possibility for patrons to engage in other activities in a more comfortable and convenient environment (Algers et al., 1975).

The total number of stops in a TSP tour includes the designated dispatch point of the bus and is given by $Q_{p(m,n)} + 1$, where $Q_{p(m,n)}$ is Poisson-distributed with a mean of $\lambda_p H_{p(m,n)} lw$. Utilizing this information along with Taylor expansion method, we derive an approximation for the expected values appearing in the right-hand-side (RHS) of Eq. (2), i.e., $E\left[\left((Q_{p(m,n)} + 1)^{\beta_3 + \frac{3}{2}} - (Q_{p(m,n)} + 1)^{\beta_3 + \frac{1}{2}}\right) e^{\beta_4(Q_{p(m,n)}+1)^{\beta_5}}\right]$ and $E[Q_{p(m,n)}^2]$. Note that *these terms are the expected values of nonlinear functions of $Q_{p(m,n)}$, and relying on the simple expected value $E[Q_{p(m,n)}]$ for their calculation could result in notable errors*. The details of the derivation of these expected values are relegated to Appendix C.

(ii) <u>In-vehicle travel time of local tours</u>

The patrons' total in-vehicle travel times in the local tour, for the outbound leg ($C_{Tp}$) and the inbound leg ($C_{Td}$) per service hour, are sums of the corresponding in-vehicle travel times for each zone, denoted $C_{Tp(m,n)}$ and $C_{Td(m,n)}$, respectively. The formulas for these components are as follows:

$$C_{Tp(m,n)} = \frac{1}{2H_{p(m,n)}} \cdot E\left[Q_{p(m,n)} \cdot \left(\frac{1}{v_I}(\beta_1 S + \beta_2)(Q_{p(m,n)} + 1)^{\beta_3} e^{\beta_4(Q_{p(m,n)}+1)^{\beta_5}} \sqrt{(Q_{p(m,n)} + 1)lw} + Q_{p(m,n)}\tau_p\right)\right]$$

$$= \frac{(\beta_1 S + \beta_2)\sqrt{lw}}{2H_{p(m,n)}v_I} E\left[\left((Q_{p(m,n)} + 1)^{\beta_3+\frac{3}{2}} - (Q_{p(m,n)} + 1)^{\beta_3+\frac{1}{2}}\right) e^{\beta_4(Q_{p(m,n)}+1)^{\beta_5}}\right] + \frac{\tau_p}{2H_{p(m,n)}} E[Q_{p(m,n)}{}^2] \quad (4)$$

$$C_{Td(m,n)} = \frac{1}{2H_{d(m,n)}} \cdot E\left[Q_{d(m,n)} \cdot \left(\frac{1}{v_I}(\beta_1 S + \beta_2)(Q_{d(m,n)} + 1)^{\beta_3} e^{\beta_4(Q_{d(m,n)}+1)^{\beta_5}} \sqrt{(Q_{d(m,n)} + 1)lw} + Q_{d(m,n)}\tau_d\right)\right]$$

$$= \frac{(\beta_1 S + \beta_2)\sqrt{lw}}{2H_{d(m,n)}v_I} E\left[\left((Q_{d(m,n)} + 1)^{\beta_3+\frac{3}{2}} - (Q_{d(m,n)} + 1)^{\beta_3+\frac{1}{2}}\right) e^{\beta_4(Q_{d(m,n)}+1)^{\beta_5}}\right] + \frac{\tau_d}{2H_{d(m,n)}} E[Q_{d(m,n)}{}^2] \quad (5)$$

$$C_{Tp} = \sum_{n=1}^{N} \sum_{m=1}^{M} C_{Tp(m,n)} \quad (6)$$

$$C_{Td} = \sum_{n=1}^{N} \sum_{m=1}^{M} C_{Td(m,n)} \quad (7)$$

where $\tau_d$ denotes the time penalty incurred at each stop in the inbound leg of service. It is important to note that, on average, a patron will experience an in-vehicle travel time that is half the total time of the vehicle's local tour, irrespective of the direction of travel. The expected values on the RHS of (4) and (5) are computed using the same method as the one employed for calculating those expected values in Eq. (2).

(iii) <u>Line-haul travel time</u>

Let $C_{Lp}$ and $C_{Ld}$ represent the total line-haul travel times for patrons in the outbound and inbound legs, respectively, per service hour. They are calculated by summing up the respective line-haul travel times for each zone $(m, n)$, denoted by $C_{Tp(m,n)}$ and $C_{Td(m,n)}$. These values are formulated as follows:

$$C_{Lp(m,n)} = \frac{1}{H_{p(m,n)}} \cdot E\left[Q_{p(m,n)} \cdot \frac{(m-1)w + (n-1)l}{v_I}\right] = \frac{(m-1)w + (n-1)l}{H_{p(m,n)}v_I} E[Q_{p(m,n)}] \quad (8)$$

$$C_{Ld(m,n)} = \frac{1}{H_{d(m,n)}} \cdot E\left[Q_{d(m,n)} \cdot \frac{(m-1)w + (n-1)l}{v_I}\right] = \frac{(m-1)w + (n-1)l}{H_{d(m,n)}v_I} E[Q_{d(m,n)}] \quad (9)$$

$$C_{Lp} = \sum_{n=1}^{N} \sum_{m=1}^{M} C_{Lp(m,n)} \quad (10)$$

$$C_{Ld} = \sum_{n=1}^{N} \sum_{m=1}^{M} C_{Ld(m,n)} \quad (11)$$

where the term $(m - 1)w + (n - 1)l$ specifies the line-haul distance for a vehicle traveling between zone $(m, n)$ and the terminal.

(iv) <u>Transfer time at the terminal</u>

Let $C_{Rp}$ and $C_{Rd}$ denote the total transfer times per service hour for patrons in the outbound and inbound directions, respectively, which are calculated as the sums of the zone-specific transfer times, denoted by $C_{Rp(m,n)}$ and $C_{Rd(m,n)}$. The formulas for these times are as follows:

$$C_{Rp(m,n)} = \frac{1}{H_{p(m,n)}} \cdot E\left[Q_{p(m,n)} \cdot \left(\frac{\tau_a Q_{p(m,n)}}{2} + t_{f-t} + \frac{H_t}{2}\right)\right] = \frac{E[Q_{p(m,n)}]}{H_{p(m,n)}}\left(t_{f-t} + \frac{H_t}{2}\right) + \frac{\tau_a}{2H_{p(m,n)}} E[Q_{p(m,n)}{}^2] \quad (12)$$

$$C_{Rd(m,n)} = \frac{1}{H_{d(m,n)}} \cdot E\left[Q_{d(m,n)} \cdot \left(\frac{\tau_b Q_{d(m,n)}}{2} + t_{t-f} + \frac{(\gamma_{(m,n)} - 1)H_{d(m,n)}}{2\gamma_{(m,n)}}\right)\right]$$

$$= \frac{E[Q_{d(m,n)}]}{H_{d(m,n)}}\left(t_{t-f} + \frac{(\gamma_{(m,n)}-1)H_{d(m,n)}}{2\gamma_{(m,n)}}\right) + \frac{\tau_b}{2H_{d(m,n)}}E[Q_{d(m,n)}{}^2] \quad (13)$$

$$C_{Rp} = \sum_{n=1}^{N}\sum_{m=1}^{M} C_{Rp(m,n)} \quad (14)$$

$$C_{Rd} = \sum_{n=1}^{N}\sum_{m=1}^{M} C_{Rd(m,n)} \quad (15)$$

where $H_t$ represents the trunk-line headway; $t_{f-t}$ and $t_{t-f}$ the outbound and inbound transfer delays per patron (including walking time), respectively; $\tau_a$ and $\tau_b$ the alighting and boarding times per patron, respectively; and $\gamma_{(m,n)}$ is an integer representing the ratio between $H_{d(m,n)}$ and $H_t$. The RHS of (12) indicates that an average patron's transfer time consists of three components: (a) the time spent in the alighting process at the rail terminal; (b) the transfer delay; and (c) the waiting time for boarding a trunk-line vehicle (e.g., a train). Eq. (13) is derived in a similar manner, with the distinction that the third component—here, the waiting time for boarding a feeder bus—is $\frac{(\gamma_{(m,n)}-1)H_{d(m,n)}}{2\gamma_{(m,n)}}$ instead of $\frac{H_t}{2}$. This difference arises from the assumptions that the feeder headway in the inbound direction, $H_{d(m,n)}$, is an integer multiple of the trunk-line headway, $H_t$, and that the feeder bus dispatches are synchronized with the trunk vehicle arrivals. This synchronization ensures that the expected load of patrons is consistent across all feeder buses.

*2.2.3 Agency cost*

The DRC service operator bears two types of costs per service hour: (i) fuel and vehicle maintenance costs, $C_{vk}$, which are proportional to the distance traveled in vehicle-kilometers; and (ii) costs associated with fleet purchase (amortized over time) and driver wages, $C_{vh}$, which are proportional to the vehicle-hours traveled. This cost structure aligns with models commonly adopted in the literature (e.g. Daganzo, 2010; Gu et al., 2016). These two costs are the sums of the corresponding cost components for each zone $(m,n)$, denoted by $C_{vkp(m,n)}$ and $C_{vhp(m,n)}$ in the outbound leg, and $C_{vkd(m,n)}$ and $C_{vhd(m,n)}$ in the inbound leg, respectively. These costs are calculated as follows:

$$C_{vkp(m,n)} = \frac{\pi_v}{\theta H_{p(m,n)}} \cdot \left((m-1)w + (n-1)l + E\left[(\beta_1 S + \beta_2)(Q_{p(m,n)} + 1)^{\beta_3} e^{\beta_4(Q_{p(m,n)}+1)^{\beta_5}}\sqrt{(Q_{p(m,n)}+1)lw}\right]\right)$$

$$= \frac{\pi_v}{\theta H_{p(m,n)}}((m-1)w + (n-1)l) + \frac{\pi_v(\beta_1 S+\beta_2)\sqrt{lw}}{\theta H_{p(m,n)}} E\left[(Q_{p(m,n)}+1)^{\beta_3+\frac{1}{2}}e^{\beta_4(Q_{p(m,n)}+1)^{\beta_5}}\right] \quad (16)$$

$$C_{vkd(m,n)} = \frac{\pi_v}{\theta H_{d(m,n)}} \cdot \left((m-1)w + (n-1)l + E\left[(\beta_1 S + \beta_2)(Q_{d(m,n)} + 1)^{\beta_3} e^{\beta_4(Q_{d(m,n)}+1)^{\beta_5}}\sqrt{(Q_{d(m,n)}+1)lw}\right]\right)$$

$$= \frac{\pi_v}{\theta H_{d(m,n)}}((m-1)w + (n-1)l) + \frac{\pi_v(\beta_1 S+\beta_2)\sqrt{lw}}{\theta H_{d(m,n)}} E\left[(Q_{d(m,n)}+1)^{\beta_3+\frac{1}{2}}e^{\beta_4(Q_{d(m,n)}+1)^{\beta_5}}\right] \quad (17)$$

$$C_{vhp(m,n)} = \frac{\pi_m}{\theta H_{p(m,n)}} \cdot \left(\frac{1}{v_I}((m-1)w + (n-1)l) + E\left[\frac{1}{v_I}(\beta_1 S + \beta_2)(Q_{p(m,n)} + \right.\right.$$

$$1)^{\beta_3}e^{\beta_4(Q_{p(m,n)}+1)^{\beta_5}}\sqrt{(Q_{p(m,n)}+1)lw}+\tau_p Q_{p(m,n)}\Bigg]\Bigg)$$

$$=\frac{\pi_m}{\theta H_{p(m,n)}v_I}\big((m-1)w+(n-1)l\big)+\frac{\pi_m(\beta_1 S+\beta_2)\sqrt{lw}}{\theta H_{p(m,n)}v_I}E\Big[(Q_{p(m,n)}+1)^{\beta_3+\frac{1}{2}}e^{\beta_4(Q_{p(m,n)}+1)^{\beta_5}}\Big]+$$
$$\frac{\pi_m\tau_p}{\theta H_{p(m,n)}}E\big[Q_{p(m,n)}\big] \tag{18}$$

$$C_{vhd(m,n)}=\frac{\pi_m}{\theta H_{d(m,n)}}\cdot\Bigg(\frac{1}{v_I}\big((m-1)w+(n-1)l\big)+E\Big[\frac{1}{v_I}(\beta_1 S+\beta_2)(Q_{d(m,n)}+$$

$$1)^{\beta_3}e^{\beta_4(Q_{d(m,n)}+1)^{\beta_5}}\sqrt{(Q_{d(m,n)}+1)lw}+\tau_d Q_{d(m,n)}\Bigg]\Bigg)$$

$$=\frac{\pi_m}{\theta H_{d(m,n)}v_I}\big((m-1)w+(n-1)l\big)+\frac{\pi_m(\beta_1 S+\beta_2)\sqrt{lw}}{\theta H_{d(m,n)}v_I}E\Big[(Q_{d(m,n)}+1)^{\beta_3+\frac{1}{2}}e^{\beta_4(Q_{d(m,n)}+1)^{\beta_5}}\Big]+$$
$$\frac{\pi_m\tau_d}{\theta H_{d(m,n)}}E\big[Q_{d(m,n)}\big] \tag{19}$$

$$C_{vk}=\sum_{n=1}^{N}\sum_{m=1}^{M}(C_{vkp(m,n)}+C_{vkd(m,n)}) \tag{20}$$

$$C_{vh}=\sum_{n=1}^{N}\sum_{m=1}^{M}(C_{vhp(m,n)}+C_{vhd(m,n)}) \tag{21}$$

where $\pi_v$ ($/vehicle-km) and $\pi_m$ ($/vehicle-hour) denote the unit costs per vehicle-km and vehicle-hour traveled, respectively; and $\theta$ ($/hour) the value of time used to convert monetary costs into equivalent temporal costs, as discussed in Section 2.1. $\pi_v$ and $\pi_m$ are expressed as linear functions of the vehicle size, $K$; see Section 4.1. Specifically, the average distance traveled by each bus, as included in the parentheses in the RHS of Eq. (16) and (17), accounts for both line-haul and local tour distances. Similarly, the average travel time per bus, factored into the RHS of Eq. (18) and (19), encompasses the cruise times for line-haul and local tours, as well as the additional time incurred at stops.

It is worth mentioning that, since asymmetric headways in the two directions are allowed, some buses would experience layovers. For simplicity, however, the costs associated with layovers and deadheading trips are disregarded. The same assumptions can be found in many previous studies (e.g., Kim and Schonfeld, 2014; Mei et al., 2021).

## 2.3 Cost models of the semi-flexible routing strategy

For ease of presentation, we will slightly abuse the notation in this section. Unless indicated otherwise, the same notations employed in the subsequent subsections retain the definitions provided in Section 2.2, with the understanding that they now refer to the context of the semi-flexible routing strategy.

### 2.3.1 Estimation of expected tour length

Consider a rectangular zone with an area $A$, in which a swath of width $w_0$ has been cut. Under the semi-flexible routing strategy, where a bus is to visit $q$ stops to pick up patrons, Daganzo (1984) provides the formula for calculating the bus's tour length using the Manhattan metric:

$$L_s=\frac{qw_0}{3}+\frac{A}{w_0} \tag{22}$$

The RHS of Eq. (22) reaches its minimum, $L_s^*=2\sqrt{\frac{qA}{3}}\approx 1.15\sqrt{qA}$, when $w_0=w_0^*=\sqrt{\frac{3A}{q}}$. However,

this minimum is unattainable if the dimensions of the zone are not integer multiples of $w_0^*$. In general, to determine the optimal swath width, the following two constraints must be met:

$$w_0 \leq \min\{l, w\} \tag{23}$$

$$w_0 \in \left\{l, w, \frac{l}{2}, \frac{w}{2}, \frac{l}{3}, \frac{w}{3}, \frac{l}{4}, \frac{w}{4}, \ldots\right\} \tag{24}$$

Constraint (23) guarantees that $w_0$ is not too large compared to the zone size. Constraint (24) ensures that the swath can be feasibly cut.

We perform numerical experiments across a range of scenarios characterized by varying numbers of stops, $q$ (ranging from 2 to 15), and aspect ratios, $S$ (ranging from 1 to 3). For each scenario, we calculate the MAPE between the minimum optimal tour length, given by $L_s^* = 1.15\sqrt{qA}$, and the tour length optimized using Eq. (22) while adhering to constraints (23) and (24). As depicted in Fig. 5, *the MAPE is substantial, reaching over 40% in scenarios with a smaller number of stops.* The results underscore the importance of considering the zone's dimensional constraints within the modeling process for semi-flexible routing.

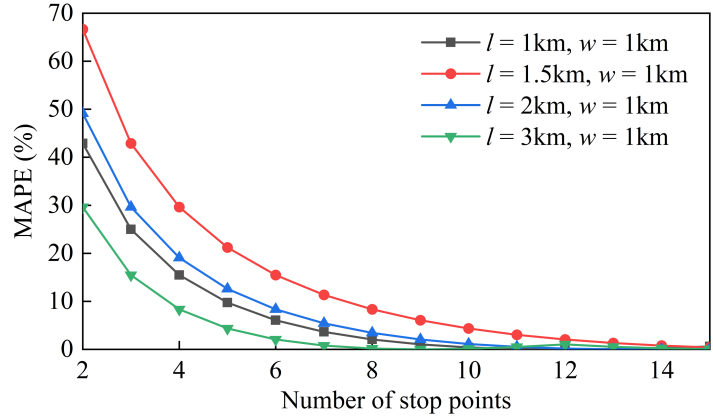

**Fig. 5. The MAPEs of tour length under semi-flexible routing**

*2.3.2 Patrons' travel time cost*

The formulations of patrons' total line-haul travel time and transfer time under semi-flexible routing are identical to those under the fully-flexible routing strategy, so we will not repeat them here. Instead, we focus on presenting the formulations for the total waiting time at home and the in-vehicle travel time during the local tour.

(i) <u>Waiting time at home</u>

$$C_{W(m,n)} = \frac{\alpha}{H_{p(m,n)}} E\left[Q_{p(m,n)}\right]\left(\frac{H_{p(m,n)}}{2} + \frac{w_0}{3v_I}\right) \tag{25}$$

$$C_W = \sum_{n=1}^{N}\sum_{m=1}^{M} C_{W(m,n)} \tag{26}$$

where $\frac{w_0}{3v_I}$ represents the average travel time for a bus to cover the lateral displacement necessary to pick up a patron, as provided by Daganzo (1984). In contrast to Eq. (2), it is evident that the waiting time under semi-flexible routing is reduced by an amount equivalent to half the local-tour travel time when compared to fully-flexible routing. This reduction arises because the waiting time with fully-flexible routing includes both the interval from the request being received to the vehicle's dispatch (which is on average half of the service headway) and the subsequent time taken for the vehicle to reach the pick-up point (the average of which is half of the local-tour travel time). In semi-flexible routing, however, the waiting time is solely composed of the period until the arrival of the next bus, which, on average, is half the service headway.

(ii) <u>In-vehicle travel time of local tours</u>

$$C_{Tp(m,n)} = \frac{1}{2H_{p(m,n)}} E\left[Q_{p(m,n)}\left(\frac{1}{v_I}\left(\frac{Q_{p(m,n)}w_0}{3} + \frac{lw}{w_0} + \frac{w_0}{2}\right) + Q_{p(m,n)}\tau_p\right)\right]$$

$$= \frac{1}{2H_{p(m,n)}}\left(\left(\frac{lw}{v_Iw_0} + \frac{w_0}{2v_I}\right)E[Q_{p(m,n)}] + \left(\frac{w_0}{3v_I} + \tau_p\right)E[Q_{p(m,n)}^2]\right) \quad (27)$$

$$C_{Td(m,n)} = \frac{1}{2H_{d(m,n)}} E\left[Q_{d(m,n)}\left(\frac{1}{v_I}\left(\frac{Q_{d(m,n)}w_0}{3} + \frac{lw}{w_0} + \frac{w_0}{2}\right) + Q_{d(m,n)}\tau_d\right)\right]$$

$$= \frac{1}{2H_{d(m,n)}}\left(\left(\frac{lw}{v_Iw_0} + \frac{w_0}{2v_I}\right)E[Q_{d(m,n)}] + \left(\frac{w_0}{3v_I} + \tau_d\right)E[Q_{d(m,n)}^2]\right) \quad (28)$$

$$C_{Tp} = \sum_{n=1}^{N}\sum_{m=1}^{M} C_{Tp(m,n)} \quad (29)$$

$$C_{Td} = \sum_{n=1}^{N}\sum_{m=1}^{M} C_{Td(m,n)} \quad (30)$$

In Eq. (27) and (28), a bus's local tour travel time is composed of cruising time and the time spent at stops. On average, a patron's in-vehicle travel time for the local tour is half the total travel time of the bus's local tour. Note that an adjustment of $\frac{w_0}{2}$ is incorporated to account for the average distance from the end of the local tour, situated (on average) along the swath's midline, to the start of the line-haul journey at the zone's bottom left corner.

*2.3.3 Agency cost*

As in Section 2.2.3, the agency costs associated with vehicle-kms and vehicle-hours traveled under the semi-flexible routing strategy are formulated as follows:

$$C_{vk(m,n)} = \frac{\pi_v}{\theta}\left(\frac{1}{H_{p(m,n)}} E\left[\frac{Q_{p(m,n)}w_0}{3} + \frac{lw}{w_0} + \frac{w_0}{2} + (m-1)w + (n-1)l\right] + \frac{1}{H_{d(m,n)}} E\left[\frac{Q_{d(m,n)}w_0}{3} + \frac{lw}{w_0} + (m-1)w + (n-1)l\right]\right)$$

$$= \frac{\pi_v}{\theta}\left(\left(\frac{1}{H_{p(m,n)}} + \frac{1}{H_{d(m,n)}}\right)\left(\frac{lw}{w_0} + \frac{w_0}{2} + (m-1)w + (n-1)l\right) + \left(\frac{E[Q_{p(m,n)}]}{H_{p(m,n)}} + \frac{E[Q_{d(m,n)}]}{H_{d(m,n)}}\right)\frac{w_0}{3}\right) \quad (31)$$

$$C_{vh(m,n)} = \frac{\pi_m}{\theta}\left(\frac{1}{H_{p(m,n)}}\left(E\left[\frac{1}{v_I}\left(\frac{Q_{p(m,n)}w_0}{3} + \frac{lw}{w_0} + \frac{w_0}{2} + (m-1)w + (n-1)l\right)\right] + E[Q_{p(m,n)}]\tau_p\right) + \frac{1}{H_{d(m,n)}}\left(E\left[\frac{1}{v_I}\left(\frac{Q_{d(m,n)}w_0}{3} + \frac{lw}{w_0} + \frac{w_0}{2} + (m-1)w + (n-1)l\right)\right] + E[Q_{d(m,n)}]\tau_d\right)\right)$$

$$= \frac{\pi_m}{\theta}\left(\frac{1}{v_I}\left(\left(\frac{1}{H_{p(m,n)}} + \frac{1}{H_{d(m,n)}}\right)\left(\frac{lw}{w_0} + \frac{w_0}{2} + (m-1)w + (n-1)l\right) + \left(\frac{E[Q_{p(m,n)}]}{H_{p(m,n)}} + \frac{E[Q_{d(m,n)}]}{H_{d(m,n)}}\right)\frac{w_0}{3}\right) + \left(\frac{E[Q_{p(m,n)}]\tau_p}{H_{p(m,n)}} + \frac{E[Q_{d(m,n)}]\tau_d}{H_{d(m,n)}}\right)\right) \quad (32)$$

$$C_{vk} = \sum_{n=1}^{N} \sum_{m=1}^{M} (C_{vk(m,n)} + C_{vk(m,n)}) \tag{33}$$

$$C_{vh} = \sum_{n=1}^{N} \sum_{m=1}^{M} (C_{vh(m,n)} + C_{vh(m,n)}) \tag{34}$$

### 2.4 Optimization model

For the fully-flexible routing strategy, the objective function, i.e., the total societal cost of the DRC system, can be written as the sum of all the cost components presented in Section 2.2:

$$GC = C_W + C_{Tp} + C_{Td} + C_{Lp} + C_{Ld} + C_{Rp} + C_{Rd} + C_{vk} + C_{vh} \tag{35a}$$

The optimization problem is formulated as follows:

$$\min GC \tag{35b}$$

subject to:

$$\lambda_p H_{p(m,n)} lw + 2\left(\lambda_p H_{p(m,n)} lw\right)^{\frac{1}{2}} \leq K \tag{35c}$$

$$\lambda_d H_{d(m,n)} lw + 2\left(\lambda_d H_{d(m,n)} lw\right)^{\frac{1}{2}} \leq K \tag{35d}$$

$$H_{d(m,n)} = \gamma_{(m,n)} H_t \tag{35e}$$

$$H_{min} \leq H_{p(m,n)} \leq H_{max} \tag{35f}$$

$$\max\{H_{min}, H_t\} \leq H_{d(m,n)} \leq H_{max} \tag{35g}$$

where $K$ denotes a DRC bus's patron-carrying capacity; $H_{min}$ and $H_{max}$ the minimum and maximum headways, respectively; and $\gamma_{(m,n)}$ is an integer. Constraints (35c) and (35d) ensure that the DRC bus's capacity is no less than the average number of patrons onboard *plus two standard deviations*. This statistically covers approximately 98% of all instances, thereby minimizing the risk of capacity violations. Constraint (35e) requires that the DRC headway in the inbound direction is an integer multiple of the trunk vehicle headway.

For the semi-flexible routing strategy, the optimization formulation remains consistent with the one used here, though the cost models differ, and it includes additional constraints (23) and (24).

## 3. Solution method

First, note that the four integer variables: $M$, $N$, $K$ and $\gamma_{(m,n)}$, as well as $w_0$ in the semi-flexible routing strategy can take values only from limited ranges. These constraints make it feasible to employ an exhaustive search to determine their optimal values. Specifically, in our approach, we define the ranges as $K \in \{1, 2, \dots, 20\}$, $M \in \{1, 2, \dots, 6\}$, $N \in \{1, 2, \dots, 6\}$, and $\gamma_{(m,n)} \in \{1, 2, \dots, 5\}$ where $m \in \{1, 2, \dots, M\}$, $n \in \{1, 2, \dots, N\}$, and $w_0 \in \left\{l, w, \frac{l}{2}, \frac{w}{2}, \frac{l}{3}, \frac{w}{3}, \frac{l}{4}, \frac{w}{4}\right\}$, subject to the condition that $w_0 \leq \min\{l, w\}$ (the values of $l$ and $w$ can be determined when $M$ and $N$ are known).

For each combination of $M$, $N$, $K$ and $\gamma_{(m,n)}$ (along with $w_0$ for semi-flexible routing), we aim to find the optimal value for $H_{p(m,n)}$ where $m \in \{1, 2, \dots, M\}$ and $n \in \{1, 2, \dots, N\}$ (note that $H_{d(m,n)}$ is determined by $\gamma_{(m,n)} H_t$, as shown in Eq. (35e)). The optimization problem for $H_{p(m,n)}$ is a nonlinear and non-convex mathematical program. We solve this problem using the sequential quadratic programming method, implemented via the "fmincon" function in Matlab R2020a. Given the non-convex nature of the problem, we execute multiple runs with different initial values to achieve a high-quality, near-optimal solution. Ultimately, we select the solution with the lowest cost from all evaluated combinations of the discrete variables.

The detailed solution algorithm under the fully-flexible routing strategy is provided below:

Step 1: Iterate over each combination of $\{M, N, K, \gamma_{(m,n)}\} \in \{1,2,...,6\} \times \{1,2,...,6\} \times \{1,2,...,20\} \times \{1,2,...,5\}$, where "×" denotes the Cartesian product:

1.1 Assign random initial values to $H_{p(m,n)}$ within the range of $[H_{min}, H_{max}]$ for $m \in \{1,2,...,M\}$ and $n \in \{1,2,...,N\}$.
1.2 Use "fmincon" to find the optimal solution for $H^*_{p(m,n)}$. Record the solution and the associated total cost.
1.3 Repeat sub-steps 1.1 and 1.2 for 20 times and record the lowest-cost solution for the current set of $\{M, N, K, \gamma_{(m,n)}\}$.[5]

Step 2: From all combinations of $\{M, N, K, \gamma_{(m,n)}\}$, select the solution with the lowest cost and output it.

Under the semi-flexible routing strategy, Step 1 is adjusted by including the feasible set of $w_0$ in the Cartesian product.

## 4. Numerical case studies

Section 4.1 delineates the setup of the numerical case studies. A comparative analysis of the performance of our models against those using the $k^*$ values/formulas from preceding research is presented in Section 4.2. The two DRC routing strategies, fully-flexible routing and semi-flexible routing, are compared in Section 4.3. Section 4.4 delves into the performance sensitivity of both DRC strategies and a fixed-route feeder design concerning various key operational parameters, including the demand density, the service region dimensions, the value of time, and the discount factor for the home-waiting time value.

### 4.1 Experimental setup

Our numerical instances utilize the parameter values outlined in Table 2, except where noted otherwise. These values are borrowed from literatures (Daganzo, 2010; Li and Quadrifoglio, 2010; Kim and Schonfeld, 2015; Sangveraphunsiri et al., 2022; Zhen and Gu, 2023). We chose a service region size of $2 \times 2$ km², which represents a quarter of a suburban center or a satellite town linked to a trunk transit line. The demand density is set at 40 patrons/h/km² for both inbound and outbound directions. A sensitivity analysis regarding these parameter values will be explored in Section 4.4.

In our models, coefficient $\alpha$ denotes the discount factor for wait time value at home. Following earlier studies (Nourbakhsh and Ouyang, 2012; Qiu et al., 2014; Zheng et al., 2018b; Petit and Ouyang, 2022), $\alpha$ varies between 0 and 1, with higher values indicating a more significant wait time impact on the overall cost. For our DRC service, $\alpha$ is set below 1, as DRC bus dispatch times are known to the patrons. This permits patrons to plan other activities during their wait, leading to a more agreeable waiting experience compared to being in transit (Ohmori et al., 2004).

---

[5] In every numerical instance we examined, different initial values consistently converged to the same solution, leading us to believe that this solution is indeed the optimal one.

The optimization problem was solved in a PC powered by an Intel Core i7-1165G7 processor with 16.0GB of RAM, clocked at 2.80GHz. The solution algorithm was implemented using Matlab 2020a. On average, solving a case required about 20 seconds of computational time.

Table 2. Parameter values

| Notation | Description | Value | Unit |
|---|---|---|---|
| $L$ | Length of the service region | 2 | km |
| $W$ | Width of the service region | 2 | km |
| $\lambda_p$ | Outbound demand density | 40 | patrons/h/km² |
| $\lambda_d$ | Inbound demand density | 40 | patrons/h/km² |
| $\theta$ | Value of time | 20 | $/h |
| $\alpha$ | Discount factor of waiting time value at home | 0.3 | - |
| $\pi_v$ | Unit cost per bus-km traveled | $0.0314 + 0.0039K$ | $/vehicle·km |
| $\pi_m$ | Unit cost per bus-hour traveled | $2.068 + 0.108K + 2\theta$ | $/vehicle·h |
| $\tau_0$ | Dwell time loss per stop due to bus acceleration and deceleration[6] | 26/3600 | h/stop |
| $\tau_p$ | Dwell time per stop in the outbound leg | 30/3600 | h/patron |
| $\tau_d$ | Dwell time per stop in the inbound leg | 28/3600 | h/patron |
| $\tau_a$ | Alighting time per patron | 2/3600 | h/patron |
| $\tau_b$ | Boarding time per patron | 4/3600 | h/patron |
| $v_I$ | Bus cruise speed | 25 | km/h |
| $t_{f-t}$ | Transfer delay from feeder to trunk transit per patron | 3/60 | h/patron |
| $t_{t-f}$ | Transfer delay from trunk transit to feeder per patron | 3/60 | h/patron |
| $H_{min}$ | Minimum feeder headway | 3/60 | h |
| $H_{max}$ | Maximum feeder headway | 1 | h |
| $H_t$ | Headway of the trunk line | 5/60 | h |
| $\beta_1$ | Coefficient of Eq. (1) | 0.1102 | - |
| $\beta_2$ | Coefficient of Eq. (1) | 1.4569 | - |
| $\beta_3$ | Coefficient of Eq. (1) | −0.1472 | - |
| $\beta_4$ | Coefficient of Eq. (1) | −2.5508 | - |
| $\beta_5$ | Coefficient of Eq. (1) | −2.6396 | - |

## 4.2 Benefits of accurate modeling

Compared to previous DRC studies, an important improvement in our work is the integration of more precise models for estimating tour lengths and the inclusion of second-order effects of stochastic demand within the framework of optimal DRC design. This section is dedicated to evaluating the impacts of these methodological enhancements.

Specifically, we conduct comparisons with select metrics against designs based on previous tour length estimates without considering the second-order effects of demand stochasticity (e.g., Chang and Schonfeld, 1991; Kim and Schonfeld, 2014). To this end, we implement Monte Carlo simulation to emulate actual costs for certain designs and contrast them with costs of our models and previous studies' models. We evaluate 32 scenarios with varying parameters: $\lambda_p = \lambda_d \in \{10,40\}$ patrons/h/km², $f \in \{0.3,0.9\}$, $\theta \in \{5,20\}$ $/h, $L \in \{2,3\}$ km, and $W \in \{2,3\}$ km. For each scenario, we use our models to optimize the DRC design under both routing strategies, then freeze the design variables ($M$, $N$, $K$, $\gamma_{(m,n)}$, $w_0$, $H_{p(m,n)}$, and $H_{d(m,n)}$) and simulate the costs. Each simulation run handles the demand randomly generated

---
[6] $\tau_0$ is only used in the fixed-route feeder optimization model (see Zhen and Gu, 2023), where it represents the bus dwell time loss per stop due to bus acceleration and deceleration, excluding boarding and alighting time losses. For fair comparison, we specify that $\tau_p = \tau_0 + \tau_b$ and $\tau_d = \tau_0 + \tau_a$.

within one hour using the spatial Poisson process. To find the expected generalized cost, we run at least 1000 simulations for each scenario and strategy. If the expected generalized cost does not converge after 1000 simulations, the number of simulation runs will be increased until convergence is achieved. We stipulate that convergence is attained when the expected generalized cost's standard deviation is less than 0.05. Then, we compare simulation results with our theoretical estimates and those using previous models. The percentage errors in total cost using various tour length estimates for fully-flexible and semi-flexible routing strategies are presented in Tables 3 and 4, respectively. Also shown are percentages of: (i) overall tour length estimation errors; (ii) errors in the cumulative passenger time loss due to pick-ups and drop-offs at stops, i.e., the aggregated values of $\sum_{m=1}^{M}\sum_{n=1}^{N}\left(\frac{\tau_p}{H_{p(m,n)}}E[Q_{p(m,n)}^2] + \frac{\tau_d}{H_{d(m,n)}}E[Q_{d(m,n)}^2]\right)$ in Eq. (4-5) across all zones; and (iii) instances where demand exceeds vehicle capacity. *We include the item (ii) to illustrate the impacts of accounting for the second-order stochastic effects of demand.*

Table 3. Modeling errors for the fully-flexible routing strategy

| Tour length estimation and second-order stochastic effect | Our model | | Tour length estimation by Yang et al. (2020) and excluding second-order stochastic effect | | Tour length estimation by Chakraborti and Chakrabarti (2000) and excluding second-order stochastic effect | |
|---|---|---|---|---|---|---|
| | Average | Maximum | Average | Maximum | Average | Maximum |
| Errors in $GC$ | 1.97% | 4.74% | 11.73% | 23.63% | 8.05% | 10.50% |
| Errors in outbound tour length | 1.32% | 4.71% | 33.05% | 68.99% | 18.19% | 25.24% |
| Errors in inbound tour length | 1.38% | 3.93% | 28.60% | 61.55% | 19.54% | 26.88% |
| Errors in cumulative pick-up time loss | 0.50% | 1.14% | 22.37% | 39.94% | 15.79% | 20.56% |
| Errors in cumulative drop-off time loss | 0.64% | 1.29% | 16.30% | 24.42% | 19.42% | 29.49% |
| Overcapacity | 0.45% | 0.57% | 13.46% | 16.93% | 12.73% | 18.02% |

Table 4. Modeling errors for the semi-flexible routing strategy

| Tour length estimation and second-order stochastic effect | Our model | | Tour length estimation using $k^* = 1.15$ and excluding second-order stochastic effect | |
|---|---|---|---|---|
| | Average | Maximum | Average | Maximum |
| Errors in $GC$ | 0.25% | 0.53% | 6.32% | 10.57% |
| Errors in outbound tour length | 0.43% | 0.72% | 36.83% | 42.81% |
| Errors in inbound tour length | 0.40% | 0.81% | 35.70% | 44.52% |
| Errors in cumulative pick-up time loss | 0.71% | 1.02% | 43.95% | 45.99% |
| Errors in cumulative drop-off time loss | 0.65% | 1.24% | 38.63% | 48.72% |
| Overcapacity | 0.43% | 0.51% | 14.45% | 17.50% |

Table 3 indicates that with fully-flexible routing, the average percentage error in total cost incorporating our $k^*$ formula and the second-order stochastic effects is under 2%, which clearly demonstrates the precision of our cost approximations. In contrast, errors exceed 8% and can reach up to 24% when using previous $k^*$ values or formulas and neglecting second-order stochastic effects. Although some readers might still deem these improvements moderate, it is because the cost components related to our model enhancements constitute a limited portion of the overall cost in the present case.

A closer look at tour length estimates reveals even greater discrepancies. Our average errors remain below 2%, whereas previous $k^*$ values or formulas result in average errors between 18-33%, with extremes up to 69%. This indicates that *for scenarios where local tour lengths significantly impact overall cost (e.g., shorter line-haul distances and larger zones, potentially due to a limited fleet size), our model enhancements will provide a substantial increase in estimation precision*.

Regarding cumulative pick-up/drop-off time losses, our model maintains average errors under 1%, while previous models that ignore second-order stochastic effects show average errors between 15-22%, with extremes reaching 40%. This suggests that *more significant improvements in generalized cost estimation are expected when pick-up/drop-off times are a larger portion of the total cost, such as in systems designed for transporting people with disabilities, who have considerably longer boarding ($\tau_p$) and alighting ($\tau_d$) times*.

Finally, our model keeps the chance of overcapacity as low as 0.45%. In comparison, overlooking second-order stochastic effects results in an overcapacity probability of 12-13%. This further underscores the necessity of considering second-order stochastic demand effects.

For the semi-flexible routing, the average total cost error of our refined model is 0.25%, marking a sizeable improvement over the previous models (6.3%). Errors in tour length estimates and cumulative pick-up/drop-off time losses are kept below 1% with our model, in contrast to the more than 30% errors produced by earlier models. Additionally, our model yields a probability of overcapacity of merely 0.43%, compared to 14% under previous models. These findings highlight the critical importance of the adjustments and improvements implemented in our modeling approach.

**4.3 Performance comparison between two routing strategies**

We conduct a head-to-head comparison between the optimized fully-flexible and semi-flexible routing strategies for the parameter values provided in Table 2. Key design variables, costs, and other metrics are presented in Table 5. Variable notations adorned with top bars represent mean values, averaged over either zones or individual patrons.

The first row of Table 5 shows a 3.1% generalized cost saving when opting for semi-flexible routing over fully-flexible routing. Upon closer examination, it becomes evident that semi-flexible routing offers both lower patron costs (row 2) and lower agency costs (row 3).

Further comparison of the patron cost components (rows 4-7) indicates that the greatest portion of savings comes from reduced waiting times for patrons (row 4). This reduction is attributed to *the increased waiting times associated with the fully-flexible routing strategy* (refer to Fig. 3a). On the other hand, *fully-flexible routing results in considerably shorter local-tour times* (row 5), primarily due to two factors: (i) the enhanced efficiency of the optimal TSP tours, evident from substantially lower $k^*$ values (rows 15-16); and (ii) each vehicle in the fully-flexible routing strategy transports fewer passengers compared to the semi-flexible strategy in the outbound direction (see row 13). The latter is caused by the shorter headway in fully-flexible routing (see row 11), a result of (i) shorter headways helping to counterbalance the negative effects of longer waiting times, and (ii) reduced round-trip times under fully-flexible routing (see row 17) enabling a higher frequency of service. Despite the efficiency gains from shorter local tours in fully-flexible routing, these do not compensate for the increased waiting times, thus leading to higher overall patron costs for this strategy in the present case study.

**Table 5. Optimal designs, costs, and other metrics under the two routing strategies**

| Row number | Parameters | Fully-flexible routing | Semi-flexible routing |
|---|---|---|---|
| 1 | $\overline{GC}$, mean total cost, min/patron | 18.29 | 17.73 |
| 2 | $\overline{UC}$, mean user cost, min/patron | 11.96 | 11.62 |
| 3 | $\overline{AC}$, mean agency cost, min/patron | 6.33 | 6.11 |
| 4 | $\bar{C}_W$, mean waiting time at home per patron, min/patron | 1.01 | 0.57 |
| 5 | $\bar{C}_T$, mean local-tour in-vehicle time, min/patron | 2.10 | 2.45 |
| 6 | $\bar{C}_L$, mean line-haul travel time, min/patron | 1.2 | 0.9 |
| 7 | $\bar{C}_R$, mean transfer time, min/patron | 2.18 | 2.19 |
| 8 | $K$, bus capacity, patrons/bus | 8 | 9 |
| 9 | $M \times N$, number of service zones partitioned | 2×2 | 1×4 |
| 10 | $w_0$, swath width under semi-flexible routing, km | - | 0.5 |
| 11 | $\bar{H}_{p(m,n)}$, mean headway per zone in the outbound direction, min | 4.98 | 6.80 |
| 12 | $\bar{H}_{d(m,n)}$, mean headway per zone in the inbound direction, min | 5.00 | 5.00 |
| 13 | $\bar{Q}_{p(m,n)}$, mean bus occupancy in the outbound direction, patrons/bus | 3.32 | 4.54 |
| 14 | $\bar{Q}_{d(m,n)}$, mean bus occupancy in the inbound direction, patrons/bus | 3.33 | 3.33 |
| 15 | $\bar{k}_p$, mean of $k^*$ in the outbound direction | 1.19 | 1.41 |
| 16 | $\bar{k}_d$, mean of $k^*$ in the inbound direction | 1.20 | 1.54 |
| 17 | Mean tour length in the outbound direction, km | 2.46 | 3.01 |
| 18 | Mean tour length in the inbound direction, km | 2.49 | 2.81 |

Furthermore, fully-flexible routing incurs higher agency costs than semi-flexible routing, largely due to its more frequent service.

An additional notable observation is the difference in zone partitioning between the two routing strategies: fully-flexible routing divides the service region into four square zones of 1×1 km², whereas semi-flexible routing uses four elongated zones of 0.5×2 km²; see row 9. *The square zoning in fully-flexible routing, which maintains a minimal aspect ratio of 1, evidently reduces the length of local tours*, as indicated by the positive coefficient $\beta_1$ in Eq. (1). Conversely, *semi-flexible routing prefers elongated zones, which naturally form narrow swaths with a small $w_0$, minimizing lateral detours for buses*.

### 4.4 Sensitivity analysis

In this section, we incorporate a fixed-route feeder bus service design (Zhen and Gu, 2023) into our comparative analysis of fully-flexible and semi-flexible routing strategies under various operating conditions. Section 4.4.1 examines how these strategies respond to changes in demand density. Sections 4.4.2 through 4.4.4 discuss the sensitivities to region size and shape, the value of time, and the discount factor for the value of home-waiting time, respectively.

*4.4.1 Sensitivity to the demand density*

In this analysis, we explore demand densities ranging from 2 to 210 patrons/h/km². The values for other parameters remain consistent with those used in Section 4.1. The outcomes are visualized in Fig. 6.

The key insight from Fig. 6a is that *the optimal service design transitions from fully-flexible routing to semi-flexible routing, and ultimately to fixed-route service as demand density increases*. At lower demand densities (below 21 patrons/h/km²), fully-flexible routing is the most cost-effective, with savings of up to 18% compared to semi-flexible routing. As demand increases, semi-flexible routing becomes the preferable option, predominantly due to its headways being more responsive to changes in demand, thereby reducing patrons' waiting times by a greater amount. Fixed-route service becomes the most advantageous option only when demand density exceeds 199 patrons/h/km², a shift attributable to its higher commercial speeds and the elimination of detours compared to DRC.

Another interesting observation comes from comparing the cost reduction rates of the three designs in Fig. 6a: as demand density grows from 2 to 210 patrons/h/km$^2$, the generalized cost per patron for the fixed-route service decreases by 52%, for the semi-flexible routing by 41%, and for the fully-flexible routing by 26%. These results indicate that the fixed-route service demonstrates the most substantial economies of scale, followed by semi-flexible routing, and then fully-flexible routing. The trend suggests that *increased routing flexibility tends to reduce economies of scale*. This finding is consistent with observations in other transport modes such as taxis and private vehicles, which offer the highest flexibility but the least economies of scale compared to structured transit systems.

For further comparison, Fig. 6a also includes the optimal DRC costs using $k^*$ values from previous studies (0.93 and 1.15). Notably, employing these historical $k^*$ values tends to significantly overestimate the benefits of DRC services and the critical demand density between DRC and fixed-route service (which is beyond the range of demand density investigated here). This underscores the importance of accurate parameter estimation in optimal service designs.

Fig. 6b highlights a notable difference in the zonal configurations adopted by fully-flexible and semi-flexible routing strategies. For fully-flexible routing, the zonal aspect ratio ($l/w$) remains between 1 and 2, showing minimal sensitivity to changes in demand density. *This indicates a preference for square zones, which help minimize local tour lengths.* In contrast, *semi-flexible routing increasingly favors more elongated zones as demand density rises*.

Figs. 6c and 6d show that for both routing strategies, as demand increases, there is a significant increase in the number of zones, while the number of patrons per vehicle only grows moderately. This pattern underscores *the effectiveness of strategic zonal division in keeping tour lengths short, thereby minimizing generalized costs*. Additionally, fully-flexible routing consistently utilizes smaller zones, resulting in fewer patrons per vehicle compared to semi-flexible routing.

*4.4.2 Sensitivity to region size and shape*

First, we explore the effects of different region sizes, ranging from 2 to 32 km$^2$, with a fixed aspect ratio of 1. Other parameters are again set to the values presented in Section 4.1. The results under the three service designs are presented in Fig. 7.

Fig. 7a shows that *fully-flexible routing offers considerable advantages in smaller service regions*, achieving cost savings of up to 18% compared to semi-flexible routing and 35% compared to fixed-route service. As the size of the region increases, semi-flexible routing becomes the preferred choice, primarily because its headways and agency costs are more sensitive to the region size, leading to significantly reduced waiting times and lower agency costs per patron. Should the region size increase beyond the scope depicted in Fig. 7a, fixed-route service is likely to become the most effective design, attributed to its higher commercial speed and reduced operating costs in larger service areas.

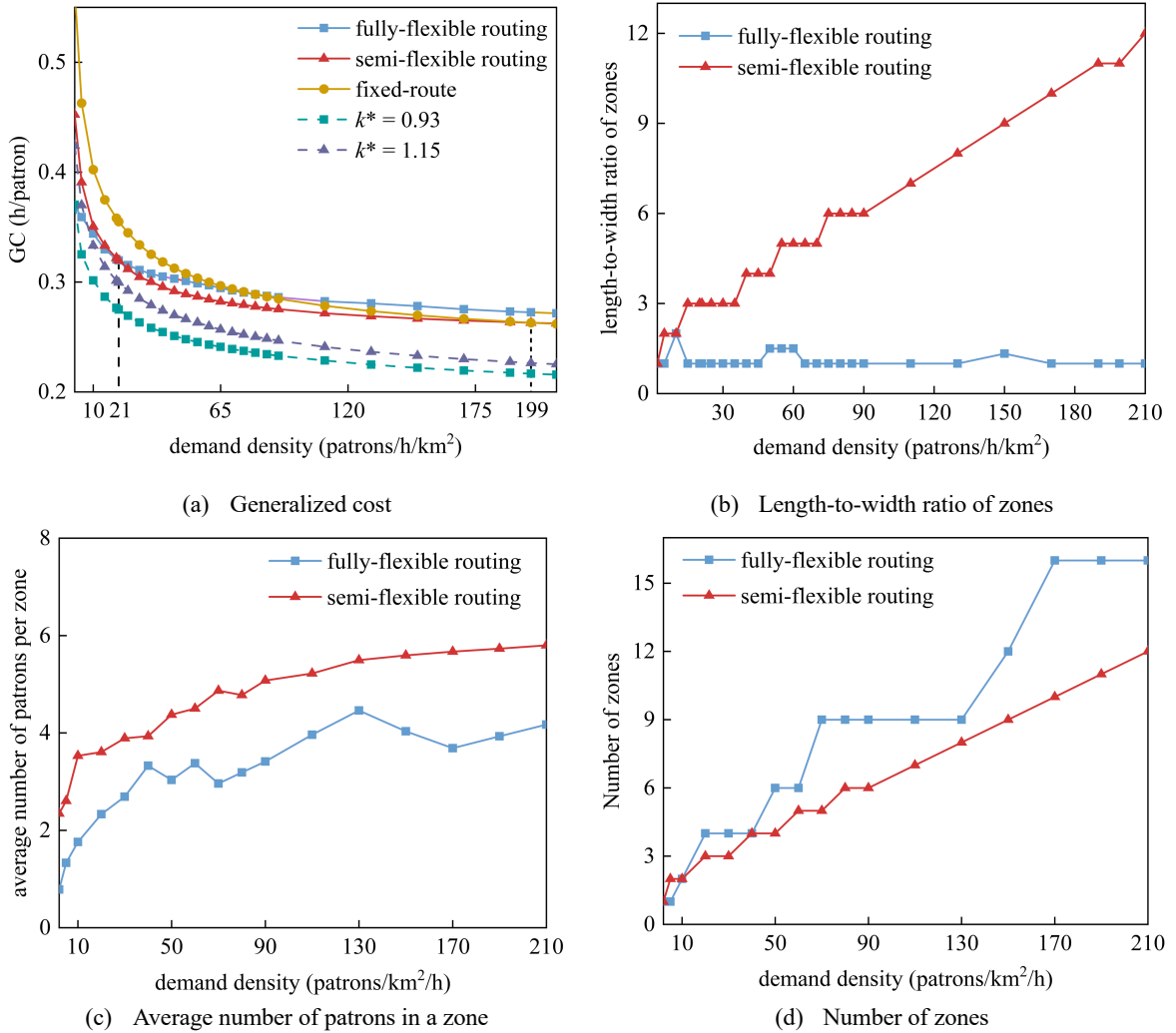

**Fig. 6. Effects of the demand density**

Fig. 6a identifies the critical demand densities at which the optimal choice of feeder service transitions among the three considered designs. Fig. 7b illustrates how these critical demand densities shift in response to changes in region size. Remarkably, these critical demand densities decrease significantly as the region size grows. Fig. 7b can be used to determine the most suitable feeder service configuration under varying operational conditions.

We also examine the impact of the service region's aspect ratio, allowing it to vary from 1 to 10 while maintaining a constant region size of 4 km². Fig. 8a shows a rising trend in generalized costs as the service region becomes more elongated, a consequence of increased tour lengths. Among the three feeder service schemes, *semi-flexible routing exhibits the least sensitivity to changes in the aspect ratio, demonstrating its superior adaptability to variations in shape*. Fig. 8b illustrates how the critical demand density threshold between fully-flexible and semi-flexible routing decreases as the aspect ratio increases, further highlighting the adaptability of semi-flexible routing to different service region shapes. The critical density distinguishing fixed-route service from DRC shows minimal sensitivity to aspect ratio changes. This curve is not depicted in Fig. 8b, as it falls outside the scope of this figure.

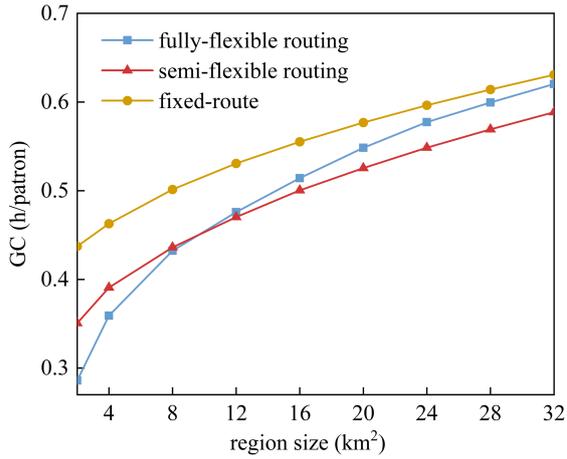
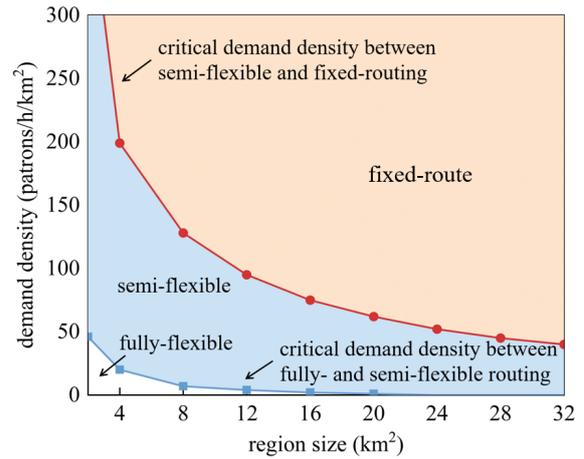

(a) Generalized costs  (b) Critical demand densities

**Fig. 7. Effects of the region size**

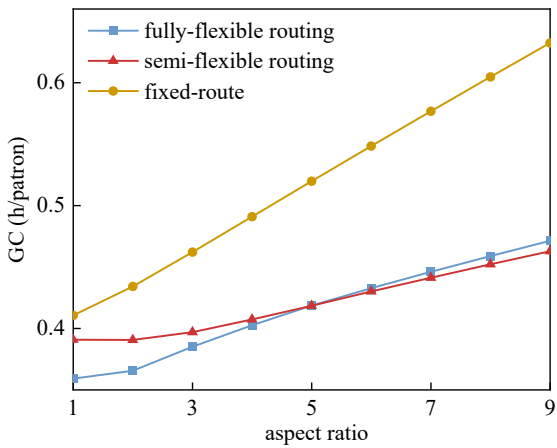
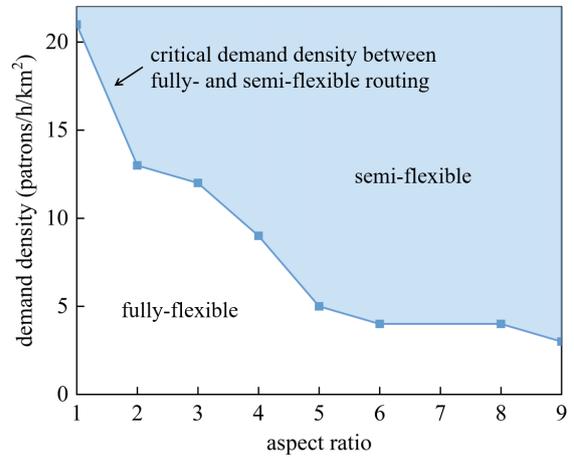

(a) Generalized costs  (b) Critical demand densities

**Fig. 8. Effects of the region shape**

*4.4.3 Sensitivity to the value of time*

Fig. 9 explores how the value of time affects the generalized costs of the three strategies, with all other parameter values held consistent with those outlined in Section 4.1. The results show that the value of time has an insignificant impact on the choice between strategies. However, as the value of time increases, the benefits of fully-flexible routing become slightly more pronounced, primarily due to a greater reduction in agency costs compared to the other alternatives.

*4.4.4 Sensitivity to the discount factor of the waiting time value at home*

Recall that the value patrons place on waiting time at home can vary significantly across different scenarios. Understanding this parameter's impact is crucial for designing effective feeder systems. Flex-route services often attribute a lower value to waiting time because they allow passengers to wait at home, offering enhanced comfort and convenience (Algers et al., 1975). This setup enables passengers to engage in other activities, thus making waiting time less burdensome and more tolerable.

In this section, we examine how the two flexible routing strategies perform under varying values of $\alpha$, which ranges from 0 to 1. Fig. 10 presents these results. For comparative purposes, results from the fixed-route service are also included, which do not depend on $\alpha$. Here we assume the value of waiting time at bus stops (for fixed-route service) is equal to that of in-vehicle travel time (Guo et al., 2011). The inbound and outbound demand densities are set at 15 patrons/km²/h.

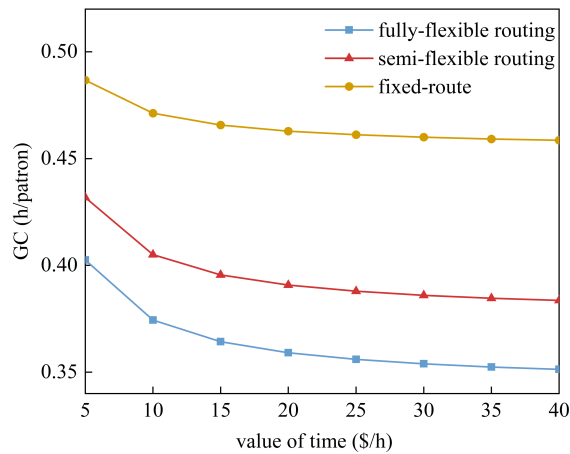

**Fig. 9. Effects of the value of time**

Fig. 10 illustrates how the effectiveness of the two strategies diverges depending on the value of $\alpha$. For $\alpha \leq 0.48$, fully-flexible routing is more advantageous, offering potential cost savings of up to 3.4%. Within this $\alpha$ range, the lower weighting of waiting time in the generalized costs compensates for the longer waiting times associated with fully-flexible routing. However, as $\alpha$ increases beyond this point, the significance of waiting time in the generalized costs grows, diminishing the benefits of fully-flexible routing due to its longer waiting times. Consequently, semi-flexible routing begins to show greater advantages, consistently outperforming fully-flexible routing by maintaining shorter waiting times for passengers. It's important to note that under the current parameter settings, fixed-route service is never the optimal choice, even with $\alpha = 1$. However, under other operating scenarios with high values of $\alpha$ (not detailed in this paper for brevity), fixed-route service could outperform the DRC alternatives.

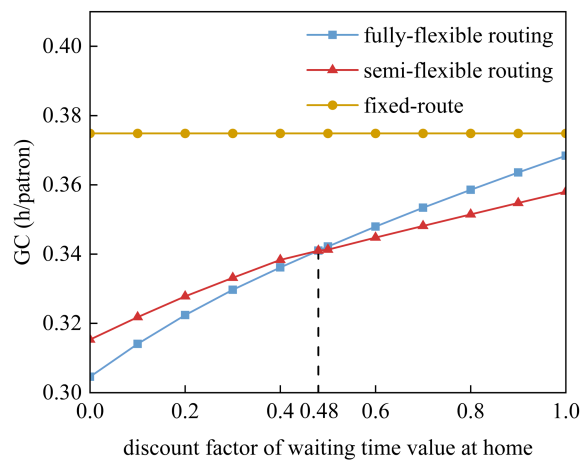

**Fig. 10. Effects of the discount factor of home-waiting time**

## 5. Conclusions

In this study, we have developed analytical models to design efficient DRC networks, focusing on two operational strategies: fully-flexible routing and semi-flexible routing. These models enable the determination of optimal zone dimensions, heterogeneous headways, and vehicle sizes, all aimed at minimizing overall system costs. We conducted extensive numerical analyses, providing a comparative evaluation of fully-flexible routing, semi-flexible routing, and fixed-route strategies across various scenarios. Our unique contributions to the literature and key findings are summarized below:

(i) To the best of our knowledge, this is the first study to clearly differentiate between two commonly used DRC routing strategies—fully-flexible and semi-flexible routing. Previous studies often conflated these strategies. We identify two main distinctions between the two strategies: whether exact optimal local tours are attained, and whether patrons are required to submit trip requests before vehicle dispatch. Consequently, while patrons served by the fully-flexible routing service benefit from shorter local-tour times, they do so at the expense of longer waiting times.

(ii) Our models significantly improve the handling of stochastic demand and the precision of tour length estimations within the optimization framework for DRC network design. By incorporating second-order effects of random passenger numbers and employing more accurate estimates of $k^*$, we achieve more precise DRC system cost estimations. For example, for fully-flexible routing, our numerical results show a reduction of the average cost estimation error from 8-12% to within 2%. For semi-flexible routing, the improvement is from 6.3% to 0.25%. Even greater improvements are expected under certain circumstances. These enhancements help rectify the previous overestimations of DRC's advantages relative to fixed-route feeder services and provide a more accurate assessment of the critical demand density threshold for choosing between these two types of services (see Fig. 6a).

(iii) The fully-flexible routing strategy proves more effective in regions with lower demand densities and smaller sizes, making it particularly suited for last-mile services. Meanwhile, the semi-flexible routing strategy excels in areas with higher demand densities and larger sizes, such as those serving airports, high-speed rail stations, and major public venues like scenic spots or hospitals.

(iv) Zoning is a critical component in the design of DRC services. The number of zones is closely aligned with demand, ensuring limited vehicle occupancy and local tour lengths. In addition, fully-flexible routing favors dividing service regions into square-shaped zones, while optimal semi-flexible routing tends towards partitioning service regions into elongated rectangular zones.

(v) As demand increases, the optimal design for feeder services evolves from fully-flexible routing to semi-flexible routing, and eventually to fixed-route services. Our findings indicate that greater routing flexibility tends to diminish the economies of scale for feeder services.

This study has several limitations, which highlight opportunities for future research. One limitation is the assumption that only one patron is picked up or dropped off per stop. Relaxing this assumption by incorporating uncertainties about the number of patrons per stop could affect calculations of patron cost components and vehicle size. Another improvement could involve combining outbound and inbound local tours, allowing the same vehicle to alternately serve pick-up and drop-off requests within a single tour. This would potentially enhance local-tour efficiency and reduce the vehicle size required.

Future research could also explore the use of modular autonomous buses for novel DRC services. Unlike the fixed-size buses considered here, modular buses can dynamically adjust their capacity based on varying demand (e.g., Liu et al., 2021; Liu et al., 2023). By adjusting the number of pods to meet demand fluctuations, modular buses can enhance both the flexibility and energy efficiency of DRC services.

Additionally, we are conducting comparative analyses between optimal flex-route feeder systems and other access modes, including shared bikes (Wu et al., 2020), fixed-route feeders (Zhen and Gu, 2023), and ride-hailing services (Fan et al., 2024). Building on this foundation, our future work aims to integrate an optimal trunk transit network (e.g., Chen et al., 2015; Sivakumaran et al., 2014; Zhen et al., 2024) with suitable feeders tailored to the specific needs of urban areas.

## Acknowledgements

This study is supported by a General Research Fund (Project No. 15224818) provided by the Research Grants Council of Hong Kong.

## Appendix A. Table of notations

**Table A1. List of notations**

| Notation | Description | Unit |
|---|---|---|
| *Decision variables* | | |
| $M$ | Number of services zones along the vertical direction | km |
| $N$ | Number of service zones along the horizontal direction | km |
| $w_0$ | Width of the swath in semi-flexible routing | km |
| $H_{p(m,n)}$ | Headway in the outbound leg of zone $(m,n)$ | h |
| $H_{d(m,n)}$ | Headway in the inbound leg of zone $(m,n)$ | h |
| $K$ | The patron-carrying capacity of a bus | patrons/bus |
| *Cost terms and parameters (all cost terms are per hour of operations)* | | |
| $C_{W(m,n)}$ | Waiting time at home of zone $(m,n)$ | h |
| $C_W$ | Total waiting time at home | h |
| $C_{Tp(m,n)}$ | Local tour travel time in the outbound leg of zone $(m,n)$ | h |
| $C_{Tp}$ | Total local tour travel time in the outbound leg | h |
| $C_{Td(m,n)}$ | Local tour travel time in the inbound leg of zone $(m,n)$ | h |
| $C_{Td}$ | Total local tour travel time in the inbound leg | h |
| $C_{Lp(m,n)}$ | Line-haul travel time in the outbound leg of zone $(m,n)$ | h |
| $C_{Lp}$ | Total line-haul travel time in the outbound leg | h |
| $C_{Ld(m,n)}$ | Line-haul travel time in the inbound leg of zone $(m,n)$ | h |
| $C_{Ld}$ | Total line-haul travel time in the inbound leg direction | h |
| $C_{Rp(m,n)}$ | Transfer time in the outbound leg of zone $(m,n)$ | h |
| $C_{Rp}$ | Total transfer time in the outbound leg | h |
| $C_{Rd(m,n)}$ | Transfer time in the inbound leg of zone $(m,n)$ | h |
| $C_{Rd}$ | Total transfer time in the inbound leg | h |
| $C_{vkp(m,n)}$ | Distance-based agency cost in the outbound leg of zone $(m,n)$ | h |
| $C_{vkd(m,n)}$ | Distance-based agency cost in the inbound leg of zone $(m,n)$ | h |
| $C_{vk}$ | Total distance-based agency cost | h |
| $C_{vhp(m,n)}$ | Fleet-based agency cost in the outbound leg of zone $(m,n)$ | h |
| $C_{vhd(m,n)}$ | Fleet-based agency cost in the inbound leg of zone $(m,n)$ | h |
| $C_{vh}$ | Total fleet-based agency cost | h |
| $GC$ | Total societal cost (generalized cost) | h |

| Symbol | Description | Units |
|---|---|---|
| $\pi_v$ | The unit cost per bus-km traveled | \$/vehicle·km |
| $\pi_m$ | The unit cost per bus hour | \$/vehicle·h |
| *Other parameters and variables* | | |
| $q$ | Number of stops in a TSP tour | - |
| $A$ | Area of the service zone | km² |
| $k^*$ | Proportionality factor in the TSP tour length formula | - |
| $S$ | Ratio of a service zone's length to width | patrons/h |
| $W$ | Width of the service region | km |
| $L$ | Length of the service region | km |
| $w$ | Width of a service zone | km |
| $l$ | Length of a service zone | km |
| $m$ | The row number of a zone (from bottom to top) | - |
| $n$ | The column number of a zone (from left to right) | - |
| $\lambda_p$ | Demand density in the outbound direction | patrons/km²/h |
| $\lambda_d$ | Demand density in the inbound direction | patrons/km²/h |
| $Q_{p(m,n)}$ | Bus occupancy in zone $(m,n)$ in the outbound leg | - |
| $Q_{d(m,n)}$ | Bus occupancy in zone $(m,n)$ in the inbound leg | - |
| $\beta_1, \beta_2, \beta_3, \beta_4, \beta_5$ | Calibrated coefficients in the optimal TSP tour length equation | - |
| $v_I$ | Bus cruising speed | km/h |
| $\tau_p$ | Dwell time per stop in the outbound leg | h |
| $\tau_d$ | Dwell time per stop in the inbound leg | h |
| $\alpha$ | Discount factor of the time value of waiting at home | - |
| $\tau_a$ | Boarding time per patron at the terminal stop | h |
| $\tau_b$ | Alighting time per patron at the terminal stop | h |
| $t_{f-t}$ | Transfer delay from the DRC stop to the rail-line platform | h/patron |
| $t_{t-f}$ | Transfer delay from the rail-line platform to the DRC stop | h/patron |
| $H_t$ | Headway of the rail service | h |
| $\gamma_{(m,n)}$ | Integer scale factor between $H_{d(m,n)}$ and $H_t$ | - |
| $\theta$ | Value of time | \$/h |
| $H_{min}$ | Minimum headway | h |
| $H_{max}$ | Maximum headway | h |

## Appendix B. Estimating $k^*$ for the optimal TSP tour length

The data set used for calibrating the formula of $k^*$ includes 56 scenarios with various combinations of $q$ and $S$ values, where $q \in \{2, 3, ..., 15\}$ and $S \in \{1, 1.5, 2, 3\}$. ($S$ is defined to be no less than 1.) Note that in practical DRC services, scenarios where $q > 15$ or $S > 3$ are uncommon.

For each scenario, we generated and solved a sufficient number of TSP instances (at least 500) using an off-the-shelf exact solver to ensure the convergence of the average $k^*$ value within a tolerance of 0.01. More information about the solver can be found at https://ww2.mathworks.cn/help/optim/ug/travelling-salesman-problem.html.

Upon analyzing the results, which are provided in Table B.1 for reference, it became evident that $k^*$ approximates a linear function of $S$. Additionally, the relationship between $k^*$ and $q$ resembles the probability density function of a Weibull distribution. Consequently, we calibrate the following formula for $k^*$:

$$k^* = (\beta_1 S + \beta_2) q^{\beta_3} e^{\beta_4 q^{\beta_5}}$$

where $\beta_1$, $\beta_2$, $\beta_3$, $\beta_4$, and $\beta_5$ are coefficients obtained through the calibration of simulated data. Eq. (1)

represents the calibrated outcome. The simulated (average) $k^*$ values are tabulated in a table below for reference.

**Table B1. Results of simulated $k^*$**

| q \ S | 2 | 3 | 4 | 5 | 6 | 7 | 8 | 9 | 10 | 11 | 12 | 13 | 14 | 15 |
|---|---|---|---|---|---|---|---|---|---|---|---|---|---|---|
| 1 | 0.94 | 1.16 | 1.20 | 1.19 | 1.18 | 1.17 | 1.15 | 1.13 | 1.12 | 1.11 | 1.10 | 1.09 | 1.09 | 1.08 |
| 1.5 | 0.96 | 1.17 | 1.22 | 1.22 | 1.20 | 1.19 | 1.17 | 1.15 | 1.15 | 1.13 | 1.12 | 1.11 | 1.10 | 1.10 |
| 2 | 1.00 | 1.23 | 1.27 | 1.28 | 1.25 | 1.23 | 1.21 | 1.20 | 1.18 | 1.16 | 1.15 | 1.14 | 1.13 | 1.12 |
| 3 | 1.09 | 1.33 | 1.38 | 1.38 | 1.36 | 1.34 | 1.31 | 1.29 | 1.27 | 1.25 | 1.23 | 1.21 | 1.20 | 1.19 |

## Appendix C. Approximations of expected values by Taylor expansion

For brevity, we use simplified notations $Q$, $\lambda$, and $H$ to represent the number of passengers a vehicle needs to pick up in a service zone, the demand density, and the headway, respectively.

Due to the assumption of spatial Poisson demand, we have:

$$E[Q] = \lambda H l w \tag{C1}$$

$$Var[Q] = E[(Q - E[Q])^2] = \lambda H l w \tag{C2}$$

$$E[Q^2] = E[Q]^2 + Var[Q] = (\lambda H l w)^2 + \lambda H l w \tag{C3}$$

where $E[\cdot]$ denotes the expected value operator and $Var[\cdot]$ the variance operator.

We derive second-order approximations for $(Q+1)^{\beta_3+\frac{3}{2}}e^{\beta_4(Q+1)^{\beta_5}}$ and $(Q+1)^{\beta_3+\frac{1}{2}}e^{\beta_4(Q+1)^{\beta_5}}$, which are used in Eq. (2), (4), (5), and (16-19). Applying Taylor expansion at the point of the expected value $E[Q]$, we have:

$$(Q+1)^{\beta_3+\frac{3}{2}}e^{\beta_4(Q+1)^{\beta_5}} = e^{\beta_4(E[Q]+1)^{\beta_5}}\left\{(E[Q]+1)^{\beta_3+\frac{3}{2}} + \left[\left(\beta_3+\frac{3}{2}\right)(E[Q]+1)^{\beta_3+\frac{1}{2}} + \beta_4\beta_5(E[Q]+1)^{\beta_3+\beta_5+\frac{1}{2}}\right](Q-E[Q]) + \left[\left(\beta_3+\frac{3}{2}\right)\left(\beta_3+\frac{1}{2}\right)(E[Q]+1)^{\beta_3-\frac{1}{2}} + \beta_4\beta_5\left(\beta_3+\frac{3}{2}\right)(E[Q]+1)^{\beta_3+\beta_5-\frac{1}{2}} + \beta_4\beta_5\left(\beta_3+\beta_5+\frac{1}{2}\right)(E[Q]+1)^{\beta_3+\beta_5-\frac{1}{2}} + \beta_4^2\beta_5^2(E[Q]+1)^{\beta_3+2\beta_5-\frac{1}{2}}\right]\frac{(Q-E[Q])^2}{2}\right\} + O((Q-E[Q])^3) \tag{C4}$$

$$(Q+1)^{\beta_3+\frac{1}{2}}e^{\beta_4(Q+1)^{\beta_5}} = e^{\beta_4(E[Q]+1)^{\beta_5}}\left\{(E[Q]+1)^{\beta_3+\frac{1}{2}} + \left[\left(\beta_3+\frac{1}{2}\right)(E[Q]+1)^{\beta_3-\frac{1}{2}} + \beta_4\beta_5(E[Q]+1)^{\beta_3+\beta_5-\frac{1}{2}}\right](Q-E[Q]) + \left[\left(\beta_3+\frac{1}{2}\right)\left(\beta_3-\frac{1}{2}\right)(E[Q]+1)^{\beta_3-\frac{3}{2}} + \beta_4\beta_5\left(\beta_3+\frac{1}{2}\right)(E[Q]+1)^{\beta_3+\beta_5-\frac{3}{2}} + \beta_4\beta_5\left(\beta_3+\beta_5-\frac{1}{2}\right)(E[Q]+1)^{\beta_3+\beta_5-\frac{3}{2}} + \beta_4^2\beta_5^2(E[Q]+1)^{\beta_3+2\beta_5-\frac{3}{2}}\right]\frac{(Q-E[Q])^2}{2}\right\} + O((Q-E[Q])^3) \tag{C5}$$

where $O(\cdot)$ indicates the big O notation.

Apply the expected value operator to both sides of (C4) and (C5) and omit the higher-order terms:

$$E\left[(Q+1)^{\beta_3+\frac{3}{2}}e^{\beta_4(Q+1)^{\beta_5}}\right] \approx e^{\beta_4(\lambda H l w+1)^{\beta_5}}\left\{(\lambda H l w+1)^{\beta_3+\frac{3}{2}} + \left[\left(\beta_3+\frac{3}{2}\right)\left(\beta_3+\frac{1}{2}\right)(\lambda H l w+1)^{\beta_3-\frac{1}{2}} + \beta_4\beta_5\left(\beta_3+\frac{3}{2}\right)(\lambda H l w+1)^{\beta_3+\beta_5-\frac{1}{2}} + \beta_4\beta_5\left(\beta_3+\beta_5+\frac{1}{2}\right)(\lambda H l w+1)^{\beta_3+\beta_5-\frac{1}{2}} + \beta_4^2\beta_5^2(\lambda H l w+1)^{\beta_3+2\beta_5-\frac{1}{2}}\right]\frac{\lambda H l w}{2}\right\} \tag{C6}$$

$$E\left[(Q+1)^{\beta_3+\frac{1}{2}}e^{\beta_4(Q+1)^{\beta_5}}\right] \approx e^{\beta_4(\lambda Hlw+1)^{\beta_5}}\left\{(\lambda Hlw+1)^{\beta_3+\frac{1}{2}} + \left[\left(\beta_3+\frac{1}{2}\right)\left(\beta_3-\frac{1}{2}\right)(\lambda Hlw+1)^{\beta_3-\frac{3}{2}} + \beta_4\beta_5\left(\beta_3+\frac{1}{2}\right)(\lambda Hlw+1)^{\beta_3+\beta_5-\frac{3}{2}} + \beta_4\beta_5\left(\beta_3+\beta_5-\frac{1}{2}\right)(\lambda Hlw+1)^{\beta_3+\beta_5-\frac{3}{2}} + \beta_4^2\beta_5^2(\lambda Hlw+1)^{\beta_3+2\beta_5-\frac{3}{2}}\right]\frac{\lambda Hlw}{2}\right\}$$
(C7)

## References

Algers, S., Hansen, S., Tegner, G., 1975. Role of waiting time, comfort, and convenience in modal choice for work trip. Transp. Res. Rec. 534(534), 38–51.

Beardwood, J., Halton, J.H., Hammersley, J.M., 1959. The shortest path through many points. Math. Proc. Cambridge Philos. Soc. 55, 299–327.

Bonomi, E., Lutton, J.L., 1984. The N-city travelling salesman problem: Statistical mechanics and the Metropolis algorithm. SIAM Rev. 26(4), 551–568.

Chakraborti, A., Chakrabarti, B.K., 2000. The travelling salesman problem on randomly diluted lattices: Results for small-size systems. Eur. Phys. J. B. 16(4), 677–680.

Chang, S.K., Schonfeld, P., 1991. Optimization models for comparing conventional and subscription bus feeder services. Transp. Sci. 25(4), 281–298.

Chen, H., Gu, W., Cassidy, M. J., Daganzo, C. F., 2015. Optimal transit service atop ring-radial and grid street networks: A continuum approximation design method and comparisons. Transport. Res. Part B 81, 755–774.

Chen, P.W., Nie, Y.M., 2017. Analysis of an idealized system of demand adaptive paired-line hybrid transit. Trans. Res. Part B 102, 38–54.

Chen, Y., Wang, H., 2018. Pricing for a last-mile transportation system. Transp. Res. Part B 107, 57–69.

Daganzo, C.F., 1984. The length of tours in zones of different shapes. Transp. Res Part B 18(2), 135–145.

Daganzo, C. F., 2009. A headway-based approach to eliminate bus bunching: Systematic analysis and comparisons. Transp. Res. Part B 43(10), 913–921.

Daganzo, C.F., 2010. Structure of competitive transit networks. Trans. Res. Part B 44(4), 434–446.

Dzisi, E.K.J., Obeng, D.A., Ackaah, W., Tuffour, Y.A., 2022. MaaS for paratransit minibus taxis in developing countries: A review. Travel Behav. Soc. 26, 18–27.

Fan, W., Gu, W., Xu, M., 2024. Optimal design of ride-pooling as on-demand feeder services. Transp. Res. Part B 185, 102964.

Fan, W., Mei, Y., Gu, W., 2018. Optimal design of intersecting bimodal transit networks in a grid city. Transp. Res. Part B 111, 203–226.

Gu, W., Amini, Z., Cassidy, M.J., 2016. Exploring alternative service schemes for busy transit corridors. Trans. Res. Part B 93, 126–145.

Guo, Q.W., Chow, J.Y., Schonfeld, P., 2017. Stochastic dynamic switching in fixed and flexible transit services as market entry-exit real options. Transp. Res. Proc. 23, 380–399.

Guo, S., Yu, L., Chen, X., Zhang, Y., 2011. Modelling waiting time for passengers transferring from rail to buses. Transp. Plan. Technol. 34(8), 795–809.

Kersting, M., Kallbach, F., Schlüter, J. C., 2021. For the young and old alike–An analysis of the determinants of seniors' satisfaction with the true door-to-door DRT system EcoBus in rural Germany. J. Transp. Geogr. 96, 103173.

Kim, M., Schonfeld, P., 2012. Conventional, flexible, and variable-type bus services. J. Transp. Eng. 138(3), 263.

Kim, M., Schonfeld, P., 2013. Integrating bus services with mixed fleets. Transp. Res. Part B 55, 227–244.

Kim, M., Schonfeld, P., 2014. Integration of conventional and flexible bus services with timed transfers. Transp. Res. Part B 68, 76–97.

Kim, M., Schonfeld, P., 2015. Maximizing net benefits for conventional and flexible bus services. Transp. Res. Part A 80, 116–133.

Kim, M., Levy, J., Schonfeld, P., 2019. Optimal zone sizes and headways for flexible-route bus services. Transp. Res. Part B 130, 67–81.

Lee, J., Choi, M., 1994. Optimization by multicanonical annealing and the traveling salesman problem. Phys. Rev. E 50(2), R651.

Li, X., Quadrifoglio, L., 2010. Feeder transit services: Choosing between fixed and demand responsive policy. Transp. Res. Part C 18(5), 770–780.

Liu, X., Qu, X., Ma, X., 2021. Improving flex-route transit services with modular autonomous vehicles. Transp. Res. Part E 149, 102331.

Liu, Y., Ouyang, Y., 2021. Mobility service design via joint optimization of transit networks and demand-responsive services. Transp. Res. Part B 151, 22–41.

Liu, Z., de Almeida Correia, G. H., Ma, Z., Li, S., Ma, X., 2023. Integrated optimization of timetable, bus formation, and vehicle scheduling in autonomous modular public transport systems. Transp. Res. Part C 155, 104306.

Ma, J., Chan, J., Ristanoski, G., Rajasegarar, S., Leckie, C., 2019. Bus travel time prediction with real-time traffic information. Transp. Res. Part C 105, 536–549.

Mei, Y., Gu, W., Cassidy, M.J., Fan, W., 2021. Planning skip-stop transit service under heterogeneous demands. Trans. Res. Part B 150, 503–523.

Montenegro, B.D.G., Sörensen, K., Vansteenwegen, P., 2021. A large neighborhood search algorithm to optimize a demand-responsive feeder service. Transp. Res. Part C 127, 103102.

Nourbakhsh, S.M., and Ouyang, Y., 2012. A structured flexible transit system for low demand areas. Transp. Res. Part B 46(1), 204–216.

Ohmori, N., Hirano, T., Harata, N., Ohta, K., 2004. Passengers' waiting behavior at bus stop. Traffic and Transportation Studies: Proceedings of ICTTS 2004, 2–4.

Ouyang, Y., Nourbakhsh, S.M., Cassidy, M.J., 2014. Continuum approximation approach to bus network design under spatially heterogeneous demand. Trans. Res. Part B 68, 333–344.

Pan, S., Yu, J., Yang, X., Liu, Y., Zou, N., 2015. Designing a flexible feeder transit system serving irregularly shaped and gated communities: Determining service area and feeder route planning. J. Urban Plan. Dev, 141(3), 04014028.

Percus, A.G., Martin, O.C., 1996. Finite size and dimensional dependence in the Euclidean traveling salesman problem. Phys. Rev. Lett. 76(8), 1188.

Petit, A., Yildirimoglu, M., Geroliminis, N., Ouyang, Y., 2021. Dedicated bus lane network design under demand diversion and dynamic traffic congestion: An aggregated network and continuous approximation model approach. Transp. Res. Part C 128, 103187.

Qiu, F., Li, W., Zhang, J., 2014. A dynamic station strategy to improve the performance of flex-route transit services. Transp. Res. Part C 48, 229–240.

Quadrifoglio, L., Li, X., 2009. A methodology to derive the critical demand density for designing and operating feeder transit services. Transp. Res. Part B 43(10), 922–935.

Sangveraphunsiri, T., Cassidy, M.J., Daganzo, C.F., 2022. Jitney-lite: a flexible-route feeder service for developing countries. Transp. Res. Part B 156, 1–13.

Stein, D.M., 1978a. An asymptotic, probabilistic analysis of a routing problem. Math. Oper. Res. 3(2), 89–101.

Stein, D.M., 1978b. Scheduling dial-a-ride transportation systems. Transp. Sci 12(3), 232–249.

Sivakumaran, K., Li, Y., Cassidy, M. J., Madanat, S., 2014. Access and the choice of transit technology. Trans. Res. Part A 46(1), 131–139.

Wang, C., Quddus, M., Enoch, M., Ryley, T., Davison, L., 2014. Multilevel modelling of Demand Responsive Transport (DRT) trips in Greater Manchester based on area-wide socio-economic data. Transportation, 41, 589–610.

Wang, J., Liu, K., Yamamoto, T., Wang, D., Lu, G, 2023. Built environment as a precondition for demand-responsive transit (DRT) system survival: Evidence from an empirical study. Travel Behav. Soc. 30, 271–280.

Wong, R.C.P., Yang, L., Szeto, W.Y., 2023. Comparing passengers' satisfaction with fixed-route and demand-responsive transport services: Empirical evidence from public light bus services in Hong Kong. Travel Behav. Soc. 32, 100583.

Wu, L., Gu, W., Fan, W., Cassidy, M. J., 2020. Optimal design of transit networks fed by shared bikes. Transp. Res. Part B 131, 63–83.

Yang, H., Liang, X., Zhang, Z., Liu, Y., Abid, M. M., 2020. Statistical modeling of quartiles, standard deviation, and buffer time index of optimal tour in traveling salesman problem and implications for travel time reliability. Transp. Res. Rec 2674(12), 339–347.

Zhen, L., Gu, W., 2023. Feeder bus service design under spatially heterogeneous demand. arXiv preprint arXiv 2309.14688.

Zhen, L., Gu, W., Zhao, X., 2024. Comparing skip-stop and all-stop transit network designs. Working paper.

Zheng, Y., Li, W., Qiu, F., 2018a. A slack arrival strategy to promote flex-route transit services. Transp. Res. Part C 92, 442–455.

Zheng, Y., Li, W., Qiu, F., 2018b. A methodology for choosing between route deviation and point deviation policies for flexible transit services. J. Adv. Transp. 6292410.

Zheng, Y., Li, W., Qiu, F., Wei, H., 2019. The benefits of introducing meeting points into flex-route transit services. Transp. Res. Part C 106, 98–112.